%% file: Zn_UVES_paper_revisions3.tex
\documentclass[useAMS,usenatbib]{mn2e}

\usepackage{subfigure}
\usepackage{graphicx}
\usepackage{amsmath}
\usepackage{verbatim}
\usepackage{slashbox}
\usepackage{multirow}
\usepackage{upgreek}

\title[Abundances in Galactic Bulge planetary nebulae]{Abundances in Galactic Bulge planetary nebulae from optical, ultraviolet and infrared observations}
\author[C. L. Smith,  A. A. Zijlstra, K. M. Gesicki and H. L. Dinerstein]{Christina L. Smith $^{1,2}$\thanks{E-mail: chrsmith@yorku.ca}, Albert A. Zijlstra$^{2,3}$, Krzysztof M. Gesicki$^{4}$, \newauthor and Harriet L. Dinerstein$^{5}$ \\ 
$^{1}$ Centre for Research in Earth and Space Sciences, York University, Toronto ON, M3J1P3, Canada \\ $^{2}$ Jodrell Bank Centre for Astrophysics, University of Manchester, Manchester, M13 9PL, UK\\ $^{3}$ Department of Physics \& Laboratory for Space Research, University of Hong Kong, Pokfulam Road, Hong Kong\\ $^{4}$ Centre for Astronomy, Faculty of Physics, Astronomy and Informatics, Nicolaus Copernicus University,\\ Grudziadzka 5, PL-87-100 Torun, Poland\\ $^{5}$ Department of Astronomy, University of Texas, 2515 Speedway, C1400, Austin, TX 78712-1205, USA} 

\begin{document}

\voffset=-0.6in

\date{2017}

\pagerange{\pageref{firstpage}--\pageref{lastpage}} \pubyear{2013?}

\maketitle

\label{firstpage}

\begin{abstract}
Iron suffers from high levels of depletion in the highly ionized environments of planetary nebulae, making the direct determination of undepleted elemental iron abundances difficult. Zinc, which does not suffer from the same depletion effects as iron, may be used as a surrogate element to measure iron abundances as there is an approximately constant zinc-to-iron ratio across a wide range of metallicities. In this paper, we report zinc abundances of six Galactic Bulge planetary nebulae determined from new observations taken with ISAAC on the Very Large Telescope, Chile, prior to the instrument's decommissioning as well as a further three based upon literature observations. UVES data of the sample planetary nebulae are presented and have been used to derive abundances, temperatures and densities of a variety of elements and ions. The abundances derived from the UVES data agree well with results from the literature. [Zn/H], determined from the ISAAC observations, is found to be generally sub-solar and [O/Zn] is found to be either consistent or enriched with respect to Solar.

\end{abstract}

\begin{keywords}

ISM: abundances; (ISM:) planetary nebulae: general; infrared: general; Galaxy: abundances; Galaxy: bulge; nuclear reactions, nucleosynthesis, abundances
\end{keywords}

\section{Introduction}


Planetary nebulae, as the brightest phase of low-to-intermediate mass stellar evolution, provide the best environments to study the detailed abundances of relatively old, low-mass stars. Planetary nebula spectra contain bright, forbidden lines which are ideal for abundance calculations. In stars, the abundance of iron relative to hydrogen is often taken as the metallicity. In planetary nebulae, iron lines are extremely weak and iron can be depleted by more than 90\%, thus O/H is taken as representing the metallicity. However, oxygen and iron have different origins and enrichment histories in galactic chemical evolution, and consequently the abundance of oxygen may not follow that of iron. [O/Fe] contains information on the star formation history: a burstlike star formation gives high [O/Fe], and constant star formation gives (over time) solar-like [O/Fe]. Thus [O/H] may not accurately represent the metallicity of planetary nebulae.

Zinc, as shown in \citet{Savage1996} for example, does not generally suffer from significant depletion. \citet{Dinerstein2001} first identified the [Zn IV] emission line in spectra of two planetary nebulae (IC 4406 and NGC 7027), subsequently using it to measure the elemental abundance of zinc. They proposed that [Zn/H] may be taken as a proxy for [Fe/H] as several studies have shown that Zn/Fe is constant at the Solar value across a wide range of metallicities, $-2<$[Fe/H]$<0$ \citep{Saito2009}. Using zinc as a tracer of metallicity simultaneously avoids the enhancements associated with oxygen and other alpha elements and the depletion effects associated with iron.

This paper presents the results of two spectral studies. The first is based around the use of new broadband optical data to determine nebular diagnostics for our sample, including electron temperatures, densities and ionic and elemental abundances. The second study builds upon the work of \citet{Smith2014b}, using new and literature \citep{Smith2014b} near-infrared spectral observations in combination with the nebular diagnostics determined from the new optical data to determine zinc abundances of our sample planetary nebulae. Application of nebular diagnostics derived from new optical data to the analysis of the near-infrared spectra provides consistency across the sample. The entire sample of nebulae for which new optical data have been obtained and all bar one of the sample with new infra-red data are Galactic Bulge planetary nebulae.

The results from the optical studies generally show constant or slight decreasing [X/Zn] with varying [Zn/H], where X is Ar, N, S, Cl or He.

The results from the infrared study show a general trend of sub-solar [Zn/H] and approximately solar [O/Zn] across the sample. There is weak anti-correlation of O/Zn with Zn/H.

\section{UVES optical data}\label{UVESdata}

\subsection{Observations}

The nebulae were observed with the Ultraviolet and Visual Echelle Spectrograph (UVES) on the ESO Very Large Telescope in 2005, under service observing program 075.D-0104. UVES consists of a blue arm and a red arm, reached via a dichroic beam splitter. The observations used Dichroic \#1, with a cross-over wavelength of 4500\AA. The blue arm with cross disperser \#2 covered the wavelength range 3260--4450\AA; the red arm with cross disperser \#3 covered 4580--6680\AA.  The instrument was used with a long slit, with slit width 0.5\arcsec. The slit length is set by the order separation to 8\arcsec \, in the blue and 11\arcsec \, in the red. The spectral resolution is approximately 60,000. The integration times were 600 seconds per spectrum and the final spectra included in this work are the result of averaging 3 such spectra, thus the total integration time per source is 1800 seconds. The data were reduced with the ESO Common Pipeline Library (CPL) (version 4.1.0), where the order merging was done manually.  Typical seeing was 1\arcsec. The spectra were extracted using IRAF\footnote{IRAF is distributed by the National Optical Astronomy Observatory, which is operated by the Association of Universities for Research in Astronomy (AURA) under a cooperative agreement with the National Science Foundation.}.

\subsection{Results}

The emission lines were identified using the line lists of the {\sc{Nebular Empirical Analysis Tool}} \citep{Wesson2012}, the Atomic Line List v2.04 (P. Van Hoof, 1999\footnote{http://www.pa.uky.edu/$\sim$peter/atomic/}) and \citet{Fang2011}. The results are presented in Table \ref{OptRes}. More than 90 different transitions were detected in the sample nebulae. Flux values are quoted in terms of H$\beta$, where the flux of H$\beta$ is 100.0. Where line identifications could not be reliably obtained, no originating transition is listed. The uncertainties quoted are those of the measurement error (including rms noise uncertainty) plus 2\% calibration uncertainties. The uncertainties introduced from the use of multiple arms is impossible to quantify.

\begin{table*}
\caption{Optical flux from UVES data, values are not corrected for interstellar extinction. Flux scale is in terms of H$\beta$ with F(H$\beta$)=100.0. Uncertain identifications are indicated with ``?" and unknown lines are indicated with ``$\dots$"}\label{OptRes}
\begin{tabular}{c c c c c c}
\hline
Wavelength & Element & $003.6+03.1$ & $004.0-03.0$ & $006.1+08.3$ & $006.4+02.0$ \\ 
\hline
3345.4 & [Ne V] & 0.55 $\pm$  0.02 & \dots  & \dots & \dots \\ 
3663.40 & H I 28-2 & \dots  & \dots  & 0.08 $\pm$  0.01 & \dots \\ 
3666.10 & H I 27-2 & \dots  & \dots  & 0.08 $\pm$  0.02 & \dots \\ 
3667.68 & H I 26-2 & \dots  & \dots  & 0.08 $\pm$  0.02 & \dots \\ 
3669.46 & H I 25-2 & \dots  & 0.21 $\pm$ 0.06 & 0.06 $\pm$  0.01 & \dots \\ 
3671.48 & H I 24-2 & \dots  & 0.26 $\pm$ 0.07 & 0.07 $\pm$  0.01 & \dots \\ 
3673.76 & H I 23-2 & \dots  & 0.23 $\pm$ 0.05 & 0.07 $\pm$  0.01 & \dots \\ 
3676.36 & H I 22-2 & \dots  & 0.22 $\pm$ 0.06 & 0.09 $\pm$  0.02 & \dots \\ 
3679.35 & H I 21-2 & 0.13 $\pm$  0.02 & 0.24 $\pm$ 0.06 & 0.1 $\pm$  0.01 & \dots \\ 
3682.81 & H I 20-2 & \dots  & 0.24 $\pm$ 0.05 & 0.13 $\pm$  0.02 & \dots \\ 
3686.83 & H I 19-2 & \dots  & 0.32 $\pm$ 0.07 & 0.09 $\pm$  0.01 & \dots \\ 
3691.55 & H I 18-2 & \dots  & 0.37 $\pm$ 0.06 & 0.14 $\pm$  0.02 & \dots \\ 
3697.15 & H I 17-2 & \dots  & 0.58 $\pm$ 0.09 & 0.13 $\pm$  0.02 & \dots \\ 
3703.85 & H I 16-2 & 0.12 $\pm$  0.02 & 0.56 $\pm$ 0.07 & 0.16 $\pm$  0.02 & \dots \\ 
3711.97 & H I 15-2 & 0.19 $\pm$  0.03 & 0.78 $\pm$ 0.1 & 0.18 $\pm$  0.02 & \dots \\ 
3721.94 & H I 14-2 & 0.28 $\pm$  0.03 & 1.06 $\pm$ 0.12 & 0.25 $\pm$  0.02 & 0.16 $\pm$  0.03 \\ 
3726.03 & [O II] & 2.48 $\pm$  0.08 & 10.44 $\pm$ 0.27 & 3.43 $\pm$  0.09 & 1.62 $\pm$  0.07 \\ 
3728.82 & [O II] & 1.00 $\pm$  0.05 & 6.06 $\pm$ 0.18 & 1.43 $\pm$  0.05 & 0.67 $\pm$  0.05 \\ 
3734.37 &  H I 13-2 & 0.16 $\pm$  0.03 & 1.05 $\pm$ 0.09 & 0.23 $\pm$  0.02 & 0.14 $\pm$  0.03 \\ 
3750.15 & H I 12-2 & 0.23 $\pm$  0.03 & 1.3 $\pm$ 0.1 & 0.31 $\pm$  0.02 & 0.19 $\pm$  0.03 \\ 
3770.63 & H I 11-2 & 0.31 $\pm$  0.03 & 1.64 $\pm$ 0.12 & 0.39 $\pm$  0.02 & 0.22 $\pm$  0.04 \\ 
3797.9 & H I 10-2 & 0.42 $\pm$  0.04 & 2.24 $\pm$ 0.14 & 0.49 $\pm$  0.03 & 0.29 $\pm$  0.04 \\ 
3819.6 & He I & 0.15 $\pm$  0.03 & 0.68 $\pm$ 0.09 & 0.1 $\pm$  0.01 & 0.06 $\pm$  0.02 \\ 
3835.38 & H I 9-2 & 0.56 $\pm$  0.04 & 3.22 $\pm$ 0.16 & 0.71 $\pm$  0.03 & 0.43 $\pm$  0.04 \\ 
3868.75 & [Ne III] & 0.88 $\pm$  0.04 & 28.78 $\pm$ 0.66 & 7.26 $\pm$  0.15 & 4.47 $\pm$  0.12 \\ 
3889.05 & H I 8-2 & 1.40 $\pm$  0.07 & 9.01 $\pm$ 0.32 & 1.78 $\pm$  0.06 & 1.13 $\pm$  0.07 \\ 
3964.73 & He I & 0.12 $\pm$  0.02 & 0.53 $\pm$ 0.08 & 0.11 $\pm$  0.01 & 0.09 $\pm$  0.02 \\ 
3967.46 & [Ne III] & 0.32 $\pm$  0.02 & 9.63 $\pm$ 0.28 & 0.97 $\pm$  0.03 & 0.13 $\pm$  0.01 \\ 
3970.07 & H I 7-2 & 1.52 $\pm$  0.06 & 1.85 $\pm$ 0.16 & 1.95 $\pm$  0.05 & 1.28 $\pm$  0.06 \\ 
4026.19 & He I & 0.36 $\pm$  0.03 & 1.34 $\pm$ 0.11 & 0.32 $\pm$  0.02 & 0.32 $\pm$  0.04 \\ 
4068.6 & [S II] & 0.39 $\pm$  0.04 & 0.25 $\pm$ 0.06 & 0.18 $\pm$  0.03 & 0.47 $\pm$  0.05 \\ 
4076.35 & [S II] & 0.15 $\pm$  0.03 & \dots  & 0.07 $\pm$  0.02 & 0.14 $\pm$  0.03 \\ 
4096 & C II & \dots  & 0.24 $\pm$ 0.1 & \dots & \dots \\ 
4101.74 & H I 6-2 & 3.48 $\pm$  0.1 & 15.54 $\pm$ 0.41 & 4.07 $\pm$  0.1 & 2.86 $\pm$  0.09 \\ 
4143.76 & He I & 0.08 $\pm$  0.02 & 0.18 $\pm$ 0.07 & 0.04 $\pm$  0.01 & \dots \\ 
4267 & C II & 0.19 $\pm$  0.03 & 0.15 $\pm$ 0.05 & 0.11 $\pm$  0.02 & \dots \\ 
4340.47 & H I 5-2 & 12.65 $\pm$  0.28 & 34.4 $\pm$ 0.79 & 14 $\pm$  0.3 & 11.47 $\pm$  0.27 \\ 
4363.21 & [O III] & 0.14 $\pm$  0.02 & 17.28 $\pm$ 0.45 & 2.21 $\pm$  0.05 & 0.57 $\pm$  0.05 \\ 
4387.92 & He I & 0.26 $\pm$  0.03 & 0.38 $\pm$ 0.08 & 0.11 $\pm$  0.02 & \dots \\ 
4471.5 & He I & 2.56 $\pm$  0.08 & 3.95 $\pm$ 0.18 & 2 $\pm$  0.05 & 2.52 $\pm$  0.09 \\ 
4630 & N II & 0.37 $\pm$  0.03 & \dots  & \dots & \dots \\ 
4633 & N III & \dots  & 0.11 $\pm$ 0.06 & \dots & \dots \\ 
4639 & O II & \dots  & 0.36 $\pm$ 0.11 & \dots & \dots \\ 
4640 & O II & \dots  & 0.22 $\pm$ 0.09 & \dots & \dots \\ 
4647 & O II & \dots  & 0.23 $\pm$ 0.07 & \dots & \dots \\ 
4711.37 & [Ar IV] & 0.79 $\pm$  0.07 & 0.66 $\pm$ 0.06 & 0.22 $\pm$  0.02 & 0.26 $\pm$  0.05 \\ 
4713.15 & He I blend & \dots  & \dots  & \dots & 0.80 $\pm$  0.08 \\ 
4740.17 & [Ar IV] & \dots  & 0.35 $\pm$ 0.05 & 0.47 $\pm$  0.02 & 0.70 $\pm$  0.07 \\ 
4810 & N II & \dots  & \dots  & 0.19 $\pm$  0.07 & \dots \\ 
4859.32 & He II & \dots  & \dots  & 0.19 $\pm$  0.07 & \dots \\ 
4861.33 & H I 4-2 & 100.00 $\pm$  2.07 & 100 $\pm$ 2.06 & 100 $\pm$  2.04 & 100.00 $\pm$  2.10 \\ 
4921.93 & He I  & \dots  & 1.32 $\pm$ 0.07 & \dots & 1.76 $\pm$  0.12 \\ 
4958.91 & [O III] & 91.86 $\pm$  1.89 & 141.81 $\pm$ 2.88 & 313.1 $\pm$  6.28 & 273.17 $\pm$  5.55 \\ 
4966 & 	[Fe VI] & \dots  & \dots  & 0.97 $\pm$  0.15 & 0.61 $\pm$  0.15 \\ 
5006.84 & [O III] & 283.66 $\pm$  5.73 & 440.02 $\pm$ 8.85 & 945.15 $\pm$  18.93 & 853.96 $\pm$  17.17 \\ 
5047.74 & He I & 0.26 $\pm$  0.05 & 0.41 $\pm$ 0.07 & 1.01 $\pm$  0.09 & 0.87 $\pm$  0.15 \\ 
5056 & 	Si II & 0.18 $\pm$  0.05 & \dots  & \dots & \dots \\ 
5197.9 & [N I] & 0.41 $\pm$  0.08 & \dots  & 0.2 $\pm$  0.05 & 0.29 $\pm$  0.07 \\ 
5200.26 & [N I] & 0.42 $\pm$  0.08 & \dots  & 0.09 $\pm$  0.03 & 0.34 $\pm$  0.08 \\ 
5270 & [Fe III] & 0.18 $\pm$  0.05 & \dots  & \dots & \dots \\ 
5517.66 & [Cl III] & 0.44 $\pm$  0.07 & 0.34 $\pm$ 0.05 & 0.24 $\pm$  0.03 & 0.53 $\pm$  0.09 \\ 
5537.6 & [Cl III] & 1.04 $\pm$  0.08 & 0.37 $\pm$ 0.04 & 0.55 $\pm$  0.04 & 1.26 $\pm$  0.12 \\ 
5577.34 & [O I] & 3.70 $\pm$  0.08 & \dots  & 7.79 $\pm$  0.16 & 25.93 $\pm$  0.53 \\ 
5679 & N II & 1.03 $\pm$  0.13 & \dots  & \dots & \dots \\ 
\hline
\end{tabular}
\end{table*}

\begin{table*}
\contcaption{}
\begin{tabular}{c c c c c c}
\hline
Wavelength & Element & $003.6+03.1$ & $004.0-03.0$ & $006.1+08.3$ & $006.4+02.0$\\ 
\hline
5711 & 	N II & 0.21 $\pm$  0.1 & \dots  & \dots & \dots \\ 
5754.6 & [N II] & 9.08 $\pm$  0.34 & 0.68 $\pm$ 0.05 & 1.89 $\pm$  0.17 & 8.94 $\pm$  0.61 \\ 
5875.66 & He I & 74.77 $\pm$  1.66 & 27.9 $\pm$ 0.63 & 44.06 $\pm$  0.95 & 89.20 $\pm$  2.13 \\ 
5941 & 	N II & 0.44 $\pm$  0.1 & \dots  & \dots & \dots \\ 
6300.3 & [O I] & 11.78 $\pm$  0.44 & 3.14 $\pm$ 0.1 & 19.08 $\pm$  0.57 & 41.09 $\pm$  1.31 \\ 
6312.1 &  [S III] & 8.30 $\pm$  0.31 & 1.89 $\pm$ 0.09 & 4.48 $\pm$  0.16 & 13.51 $\pm$  0.63 \\ 
6347 & 	Si II & 1.04 $\pm$  0.18 & \dots  & 0.24 $\pm$  0.09 & 0.82 $\pm$  0.26 \\ 
6363.77 & [O I] & 4.45 $\pm$  0.29 & 1.07 $\pm$ 0.06 & 6.8 $\pm$  0.32 & 15.54 $\pm$  0.80 \\ 
6371 & 	Si II & 0.53 $\pm$  0.13 & \dots  & 0.12 $\pm$  0.06 & 0.63 $\pm$  0.30 \\ 
6402 & 	N II? & 0.38 $\pm$  0.12 & \dots  & \dots & \dots \\ 
6481 & [Fe II], N II? & 0.63 $\pm$  0.15 & \dots  & \dots & \dots \\ 
6548.1 & [N II] & 735.04 $\pm$  14.9 & 14.63 $\pm$ 0.34 & 55.21 $\pm$  1.26 & 405.52 $\pm$  8.58 \\ 
6562.77 & H I 3-2 & 2469.76 $\pm$  49.6 & 596.58 $\pm$ 12.02 & 1713.6 $\pm$  34.38 & 3062.78 $\pm$  61.65 \\ 
6583.5 & [N II] & 2300.16 $\pm$  46.2 & 49.19 $\pm$ 1.03 & 173.81 $\pm$  3.63 & 1277.19 $\pm$  26.02 \\ 
6606 & 	[Fe II] & 0.33 $\pm$  0.08 & \dots  & \dots & \dots \\ 
\hline
\hline
Wavelength & Element & $006.8+04.1$ & $354.5+03.3$ & $355.4-02.4$ & $355.9+03.6$ \\
\hline
3682.81 & H I 20-2 & \dots & \dots & \dots & 0.18 $\pm$  0.04 \\ 
3686.83 & H I 19-2 & \dots & \dots & \dots & 0.07 $\pm$  0.02 \\ 
3691.55 & H I 18-2 & \dots & \dots & \dots & 0.13 $\pm$  0.03 \\ 
3697.15 & H I 17-2 & \dots & \dots & \dots & 0.19 $\pm$  0.05 \\ 
3703.85 & H I 16-2 & \dots & \dots & \dots & 0.11 $\pm$  0.03 \\ 
3711.97 & H I 15-2 & \dots & \dots & \dots & 0.21 $\pm$  0.11 \\ 
3721.94 & H I 14-2 & 0.19 $\pm$ 0.04 & 0.28 $\pm$ 0.04 & 0.28 $\pm$  0.04 & 0.34 $\pm$  0.07 \\ 
3726.03 & [O II] & 1.44 $\pm$ 0.07 & 1.88 $\pm$ 0.09 & 1.88 $\pm$  0.09 & 3.64 $\pm$  0.12 \\ 
3728.82 & [O II] & 0.66 $\pm$ 0.06 & 0.9 $\pm$ 0.07 & 0.9 $\pm$  0.07 & 1.29 $\pm$  0.07 \\ 
3734.37 &  H I 13-2 & 0.16 $\pm$ 0.03 & 0.19 $\pm$ 0.05 & 0.19 $\pm$  0.05 & 0.21 $\pm$  0.05 \\ 
3750.15 & H I 12-2 & 0.26 $\pm$ 0.04 & 0.22 $\pm$ 0.05 & 0.22 $\pm$  0.05 & 0.23 $\pm$  0.05 \\ 
3770.63 & H I 11-2 & 0.25 $\pm$ 0.05 & 0.29 $\pm$ 0.05 & 0.29 $\pm$  0.05 & 0.28 $\pm$  0.04 \\ 
3797.9 & H I 10-2 & 0.29 $\pm$ 0.05 & 0.39 $\pm$ 0.06 & 0.39 $\pm$  0.06 & 0.39 $\pm$  0.05 \\ 
3835.38 & H I 9-2 & 0.46 $\pm$ 0.05 & 0.44 $\pm$ 0.06 & 0.44 $\pm$  0.06 & 0.49 $\pm$  0.06 \\ 
3868.75 & [Ne III] & 4.82 $\pm$ 0.15 & 7.55 $\pm$ 0.2 & 7.55 $\pm$  0.2 & 0.19 $\pm$  0.04 \\ 
3889.05 & H I 8-2 & 1.12 $\pm$ 0.09 & 1.46 $\pm$ 0.1 & 1.46 $\pm$  0.1 & 1.06 $\pm$  0.08 \\ 
3967.46 & [Ne III] & 1.02 $\pm$ 0.06 & 2.18 $\pm$ 0.09 & 2.18 $\pm$  0.09 & 0.07 $\pm$  0.02 \\ 
3970.07 & H I 7-2 & 1.16 $\pm$ 0.08 & 1.12 $\pm$ 0.08 & 1.12 $\pm$  0.08 & 0.72 $\pm$  0.05 \\ 
4026.19 & He I & 0.26 $\pm$ 0.06 & 0.34 $\pm$ 0.06 & 0.34 $\pm$  0.06 & 0.18 $\pm$  0.04 \\ 
4068.6 & [S II] & 0.21 $\pm$ 0.04 & 0.48 $\pm$ 0.05 & 0.48 $\pm$  0.05 & 0.31 $\pm$  0.05 \\ 
4076.35 & [S II] & 0.10 $\pm$ 0.03 & 0.17 $\pm$ 0.04 & 0.17 $\pm$  0.04 & 0.1 $\pm$  0.03 \\ 
4096 & C II & \dots & 0.26 $\pm$ 0.06 & 0.26 $\pm$  0.06 & \dots \\ 
4101.74 & H I 6-2 & 2.85 $\pm$ 0.12 & 3.06 $\pm$ 0.13 & 3.06 $\pm$  0.13 & 3.26 $\pm$  0.12 \\ 
4267 & C II & \dots & 0.28 $\pm$ 0.05 & 0.28 $\pm$  0.05 & \dots \\ 
4340.47 & H I 5-2 & 11.00 $\pm$ 0.29 & 11.7 $\pm$ 0.31 & 11.7 $\pm$  0.31 & 12.16 $\pm$  0.3 \\ 
4363.21 & [O III] & 0.95 $\pm$ 0.07 & 0.96 $\pm$ 0.06 & 0.96 $\pm$  0.06 & 0.43 $\pm$  0.05 \\ 
4387.92 & He I & \dots & 0.43 $\pm$ 0.04 & 0.43 $\pm$  0.04 & \dots \\ 
4471.5 & He I & 2.13 $\pm$ 0.10 & 2.44 $\pm$ 0.11 & 2.44 $\pm$  0.11 & 1.27 $\pm$  0.07 \\ 
4633 & N III & \dots & 1.6 $\pm$ 0.23 & 1.6 $\pm$  0.23 & \dots \\ 
4639 & O II & \dots & 3.07 $\pm$ 0.28 & 3.07 $\pm$  0.28 & \dots \\ 
4640 & O II & \dots & 0.95 $\pm$ 0.17 & 0.95 $\pm$  0.17 & \dots \\ 
4656 & O II & \dots & \dots & \dots & 1.93 $\pm$  0.15 \\ 
4685.68 & He II & \dots & 16.42 $\pm$ 0.58 & 16.42 $\pm$  0.58 & \dots \\ 
4699 & [Fe III] & \dots & \dots & \dots & 0.69 $\pm$  0.12 \\ 
4711.37 & [Ar IV] & 0.60 $\pm$ 0.14 & 2.26 $\pm$ 0.29 & 2.26 $\pm$  0.29 & 0.47 $\pm$  0.07 \\ 
4713.15 & He I blend & 0.55 $\pm$ 0.09 & 0.77 $\pm$ 0.17 & 0.77 $\pm$  0.17 & \dots \\ 
4732 & 	[Fe III] & \dots & \dots & \dots & 0.37 $\pm$  0.08 \\ 
4740.17 & [Ar IV] & 0.80 $\pm$ 0.14 & 3.57 $\pm$ 0.28 & 3.57 $\pm$  0.28 & \dots \\ 
4753 & [Fe III] & \dots & \dots & \dots & 0.3 $\pm$  0.06 \\ 
4859.32 & He II & \dots & 0.82 $\pm$ 0.23 & 0.82 $\pm$  0.23 & \dots \\ 
4861.33 & H I 4-2 & 100.00 $\pm$ 2.20 & 100 $\pm$ 2.3 & 100 $\pm$  2.3 & 100 $\pm$  2.14 \\ 
4879 & [Fe III] & \dots & \dots & \dots & 0.61 $\pm$  0.12 \\ 
4921.93 & He I  & 1.42 $\pm$ 0.18 & 1.45 $\pm$ 0.26 & 1.45 $\pm$  0.26 & 0.74 $\pm$  0.1 \\ 
4958.91 & [O III] & 317.73 $\pm$ 6.52 & 325.45 $\pm$ 6.76 & 325.45 $\pm$  6.76 & 64.3 $\pm$  1.39 \\ 
5006.84 & [O III] & 993.27 $\pm$ 20.03 & 1005.22 $\pm$ 20.36 & 1005.22 $\pm$  20.36 & 199.58 $\pm$  4.09 \\ 
5010 & \dots & \dots & \dots & \dots & 0.42 $\pm$  0.1 \\ 
\hline
\end{tabular}
\end{table*}

\begin{table*}
\contcaption{}
\begin{tabular}{c c c c c c}
\hline
Wavelength & Element & $006.8+04.1$ & $354.5+03.3$ & $355.4-02.4$ & $355.9+03.6$ \\
\hline
5015.68 & He I & \dots & 2.86 $\pm$ 0.3 & 2.86 $\pm$  0.3 & \dots \\ 
5040 & 	Si II & \dots & 1.01 $\pm$ 0.27 & 1.01 $\pm$  0.27 & \dots \\ 
5047.74 & He I & 0.94 $\pm$ 0.19 & 1.61 $\pm$ 0.52 & 1.61 $\pm$  0.52 & 0.31 $\pm$  0.12 \\ 
5056 & Si II & \dots & 0.5 $\pm$ 0.23 & 0.5 $\pm$  0.23 & 0.3 $\pm$  0.09 \\ 
5197.9 & [N I] & \dots & 1.27 $\pm$ 0.27 & 1.27 $\pm$  0.27 & \dots \\ 
5200.26 & [N I] & \dots & 1.11 $\pm$ 0.27 & 1.11 $\pm$  0.27 & \dots \\ 
5270 & [Fe III] & \dots & \dots & \dots & 1.59 $\pm$  0.16 \\ 
5297 & \dots & \dots & \dots & \dots & 0.24 $\pm$  0.09 \\ 
5411.53 & He II & \dots & 2.1 $\pm$ 0.28 & 2.1 $\pm$  0.28 & \dots \\ 
5517.66 & [Cl III] & 0.70 $\pm$ 0.14 & 0.86 $\pm$ 0.26 & 0.86 $\pm$  0.26 & \dots \\ 
5537.6 & [Cl III] & 1.17 $\pm$ 0.18 & 1.7 $\pm$ 0.27 & 1.7 $\pm$  0.27 & 0.44 $\pm$  0.11 \\ 
5754.6 & [N II] & 3.11 $\pm$ 0.46 & 8.6 $\pm$ 0.6 & 8.6 $\pm$  0.6 & 12.26 $\pm$  0.57 \\ 
5875.66 & He I & 71.50 $\pm$ 1.97 & 74.7 $\pm$ 2.16 & 74.7 $\pm$  2.16 & 41.74 $\pm$  1.18 \\ 
6045 & \dots & \dots & \dots & \dots & 1.07 $\pm$  0.4 \\ 
6101 & 	[K IV] & \dots & 0.96 $\pm$ 0.42 & 0.96 $\pm$  0.42 & \dots \\ 
6300.3 & [O I] & 20.82 $\pm$ 0.90 & 41.92 $\pm$ 1.52 & 41.92 $\pm$  1.52 & 10.31 $\pm$  0.63 \\ 
6312.1 &  [S III] & 12.45 $\pm$ 0.72 & 14.58 $\pm$ 0.86 & 14.58 $\pm$  0.86 & 15.3 $\pm$  0.65 \\ 
6363.77 & [O I] & 8.09 $\pm$ 0.67 & 15.19 $\pm$ 0.95 & 15.19 $\pm$  0.95 & 3.9 $\pm$  0.52 \\ 
6371 & Si II & \dots & 2.37 $\pm$ 0.56 & 2.37 $\pm$  0.56 & \dots \\ 
6548.1 & [N II] & 161.95 $\pm$ 3.74 & 710.88 $\pm$ 14.8 & 710.88 $\pm$  14.8 & 187.82 $\pm$  4.15 \\ 
6562.77 & H I 3-2 & 3144.09 $\pm$ 63.52 & 2504.8 $\pm$ 50.87 & 2504.8 $\pm$  50.87 & 2674.47 $\pm$  53.94 \\ 
6583.5 & [N II] & 505.40 $\pm$ 10.62 & 2131.19 $\pm$ 43.33 & 2131.19 $\pm$  43.33 & 598.19 $\pm$  12.36 \\ 
\hline
\end{tabular}
\end{table*}

\begin{table*}
\caption{Electron temperatures and densities for the UVES sample of planetary nebulae. Electron temperatures are given in units of $10^3$ K and electron densities are given in units of $10^3$ cm$^{-3}$. The planetary nebula designation is given in column 1, the results from this paper (abbreviated to ``T.P.") are shown in columns 2, 4, 6, 8, \& 10. Literature values are given in columns 3, 5, 7, 9, 11 \& 12 and their origins are shown in column 13.}\label{TeRes}
{\renewcommand{\arraystretch}{1.5}
\begin{tabular}{c c c c c c c c c c c c c }
\hline
\multirow{2}{*}{PNG} & \multicolumn{2}{c}{T$_\text{e}$(N II)} & \multicolumn{2}{c}{T$_\text{e}$(O III)} & \multicolumn{2}{c}{N$_\text{e}$(Ar IV)} & \multicolumn{2}{c}{N$_\text{e}$(Cl III)} & \multicolumn{2}{c}{N$_\text{e}$(O II)}  & N$_\text{e}$(S II)  & \multirow{2}{*}{Ref.} \\
	& T.P. & Lit. & T.P. & Lit. & T.P. & Lit. & T.P. & Lit. & T.P. & Lit. & Lit. &\\
\hline
$003.6+03.1$ &  $7.7^{+0.3}_{-0.3}$ &  $8.1\pm0.1$ &  $6.8^{+0.2}_{-0.2}$  & $7.9^{+0.7}_{-0.9}$ &  \dots  & \dots	&$11^{+5}_{-0}$	&  \dots & $9^{+4}_{-0}$  & \dots &  $11\pm1$ & 1 \\ 
$004.0-03.0$ &  $10.9^{+0.4}_{-0.4}$  	& \dots 		& \dots  		& $19\pm3$ 		&  \dots  	& \dots &$2^{+1}_{-2}$  & \dots &$3.0^{+0.3}_{-0.4}$ 	& $1.6^{+1.6}_{-0.8}$	& $3^{+3}_{-2}$ & 2 \\ 
$006.1+08.3$ &  $10.7^{+0.6}_{-0.5}$  	& $11.2\pm{0.5}$ 			&  $9.5^{+0.1}_{-0.9}$  & $9.9\pm{0.4}$ 			&$16^{+3}_{-3}$	& \dots &$13^{+5}_{-5}$ & \dots &  $10^{+2}_{-0}$	& $10\pm3$ 		& $10\pm3$ & 3 \\ 
$006.4+02.0$ &  $9.9^{+0.9}_{-0.6}$  	&  $10.7^{+0.7}_{-3.5}$	& $7.4^{+0.2}_{-0.1}$  	&$7.6^{+0.3}_{-0.4}$	&$20^{+8}_{-0}$ & \dots &$11^{+5}_{-0}$ & \dots & $7^{+4}_{-0}$ 	& \dots & $9^{+5}_{-3}$   & 4 \\ 
$006.8+04.1$ &  $9.6^{+0.6}_{-0.8}$  	&  $10.6^{+0.4}_{-0.5}$	& $8.1^{+0.1}_{-0.2}$  	& $8.4\pm{2}$		&  \dots  	& \dots & $6^{+4}_{-0}$ & \dots	&  $5^{+2}_{-0}$  	&\dots & $5^{+3}_{-1}$  & 4 \\
$354.5+03.3$ &  $16^{+2}_{-3}$  	& 	$16^{+4}_{-3} $		&  $10.6^{+0.6}_{-0.5}$	& 	\dots		&  \dots  	& \dots 	& \dots  	&  \dots	& $5\pm{3}$  		&	\dots	& $17^{+1}_{-2}$ & 1\\ 
$355.4-02.4$ &  $8.1^{+0.2}_{-0.3}$  	& $8.2\pm0.4$ &  $7.8^{+0.1}_{-0.2}$ 	& 	$8.5\pm{1.0}$	&  $8^{+3}_{-3}$&  	& $11^{+9}_{-0}$& 	&  $4^{+1}_{-0}$  	&		& $3.3\pm0.4$ & 5 \\ 
$355.9+03.6$ &  $9.5^{+3}_{-0.4}$  	& 	$11^{+7}_{-1}$		&  $10.0^{+0.3}_{-0.4}$	& 	$11\pm1$		&  \dots 	& \dots	&  \dots   	& \dots	&  $30^{+60}_{-0}$   	&	\dots	& $100^{+0}_{-73}$ & 4\\ 
\hline
\end{tabular}
}

{1: \citet{Gorny2004}, 2: \citet{Exter2004}, 3: \citet{Wang2007}, 4: \citet{Gorny2009}, 5: \citet{Cuisinier2000} }
\end{table*}

\begin{table*}
\caption{Elemental and ionic abundance ratios for the UVES sample of planetary nebulae. The extinction coefficients are calculated using the H$\alpha$, H$\beta$ and H$\gamma$ line ratios \citep{Wesson2012}.}\label{abns}
{\renewcommand{\arraystretch}{1.4}
\begin{tabular}{c c c c c c c c c}
\hline
\multirow{2}{*}{PNG}  &  Ar$^{3+}$/H  &  Ar/H  &  Cl$^{2+}$/H  &  Cl/H  &  N$^{+}$/H  &  N/H  &  Ne$^{2+}$/H  &  Ne$^{4+}$/H  \\ 
  &  ($\times10^{-7}$)  &  ($\times10^{-7}$)  &  ($\times10^{-7}$)  &  ($\times10^{-7}$)  &  ($\times10^{-5}$)  &  ($\times10^{-4}$)  &   ($\times10^{-5}$) &  ($\times10^{-4}$)  \\ 
\hline
$003.6+03.1$   &  \dots   &  \dots   &  $1.6^{+0.3}_{-0.5}$   &  $2.3^{+0.5}_{-0.5}$  &  $11^{+1}_{-0}$   &  $7^{+2}_{-1}$   &  $2.8^{+0.5}_{-0.0}$ &  $1.6^{+0.3}_{-0.4}$ \\ 
$004.0-03.0$   &  $1.2^{+0.2}_{-0.3}$   &  $1.4^{+0.2}_{-0.3}$   &  $0.25^{+0.03}_{-0.04}$ &  $0.41^{+0.05}_{-0.06}$ &  $0.32^{+0.03}_{-0.03}$ &  $0.31^{+0.02}_{-0.03}$ &  $4.0^{+0.6}_{-0.6}$ &  \dots  \\ 
$006.1+08.3$   &  $1.60^{+0.09}_{-0.1}$ &  $1.7^{+0.1}_{-0.1}$ &  $0.33^{+0.03}_{-0.03}$ &  $0.66^{+0.08}_{-0.12}$ &  $0.44^{+0.06}_{-0.08}$ &  $1.1^{+0.2}_{-0.2}$ &  $2.9^{+0.2}_{-0.1}$ &  \dots  \\ 
$006.4+02.0$   &  $6.7^{+0.9}_{-0.9}$   &  $6.7^{+0.9}_{-1.2}$   &  $1.5^{+0.2}_{-0.2}$   &  $4^{+1}_{-2}$  &  $2.3^{+0.7}_{-0.4}$  &  $11.2^{+4}_{-6}$  &  $11^{+1}_{-1}$ &  \dots  \\ 
$006.8+04.1$   &  \dots   &  \dots   &  $0.8^{+0.1}_{-0.2}$  &  $2.1^{+0.5}_{-0.6}$  &  $0.9^{+0.2}_{-0.2}$  &  $4^{+1}_{-1}$  &  $6.9^{+0.7}_{-0.7}$ &  \dots \\ 
$354.5+03.3$   &  \dots   &  \dots  &  \dots   &  \dots   &  $0.4\pm0.1$  &  $6\pm2$   &  $5\pm{1}$ &  \dots \\ 
$355.4-02.4$ &  $28^{+3}_{-5}$  &  $32^{+4}_{-5}$  &  $1.5^{+0.3}_{-0.4}$   &  $3.4^{+0.9}_{-0.8}$  &  $7.9^{+1.0}_{-0.9}$  &  $24^{+6}_{-4}$  &  $11^{+1}_{-1}$ &  \dots  \\ 
$355.9+03.6$   &  \dots   &  \dots   &  $0.17^{+0.04}_{-0.04}$  &  $0.18^{+0.04}_{-0.06}$  &  $3^{+1E12}_{-2}$  &  $0.36^{+0.05}_{-0.07}$ &  $0.10^{+0.02}_{-0.02}$ &  \dots \\ 
\hline 
\end{tabular}
\begin{tabular}{c c c c c c c c c}
\hline
\multirow{2}{*}{PNG}  &  He/H  &  O$^+$/H  &  O$^{2+}$/H  &  O/H  &  S$^+$/H &  S$^{++}$/H &  S/H   & \multirow{2}{*}{$c(H\beta)$} \\ 
  &  ($\times10^{-1}$)  &   ($\times10^{-5}$)  &  ($\times10^{-4}$)  &  ($\times10^{-4}$)  &  ($\times10^{-7}$)  &  ($\times10^{-5}$)  &    ($\times10^{-5}$)  & \\ 
\hline &  &  &  &  &  &  &  \\ 
$003.6+03.1$   &  $1.12^{+0.02}_{-0.03}$   &  $6^{+3}_{-0}$ &  $3.5^{+0.5}_{-0}$ &  $4.4^{+0.8}_{-1.1}$ &  $7.0^{+1.1}_{-0.9}$ &  $1.8^{+0.4}_{-0.4}$ &  $2.4^{+0.6}_{-0.6}$ & $2.90\pm0.04$\\ 
$004.0-03.0$   &  $1.13^{+0.04}_{-0.04}$   &  $1.1^{+0.2}_{-0.2}$  &  $1.1^{+0.1}_{-0.1}$ &  $1.2^{+0.1}_{-0.1}$ &  $0.6^{+0.1}_{-0.2}$ &  $0.15^{+0.02}_{-0.03}$ &  $0.25^{+0.04}_{-0.04}$ &  $1.01\pm0.03$ \\
$006.1+08.3$   &  $0.81^{+0.02}_{-0.02}$   &  $1.5^{+0.5}_{-0.4}$ &  $3.4^{+0.1}_{-0.2}$ &  $3.6^{+0.2}_{-0.2}$ &  $0.7^{+0.2}_{-0.1}$ &  $0.25^{+0.02}_{-0.01}$ &  $0.51^{+0.06}_{-0.06}$ & $2.45\pm0.04$ \\ 
$006.4+02.0$   &  $1.17^{+0.03}_{-0.03}$   &  $1.2^{+0.7}_{-0}$ &  $7.9^{+0.8}_{-0.7}$ &  $8.1^{+1.0}_{-0.8}$ &  $3.3^{+0.6}_{-0.8}$ &  $1.5^{+0.2}_{-0}$ &  $4^{+1}_{-1}$ & $3.20\pm0.04$ \\ 
$006.8+04.1$   &  $0.94^{+0.03}_{-0.03}$   &  $0.9^{+0.4}_{-0}$ &  $5.8^{+0.5}_{-0.5}$ &  $5.8^{+0.5}_{-0.7}$ &  $1.7^{+0.5}_{-0}$ &  $0.83^{+0.09}_{-0.13}$ &  $2.2^{+0.5}_{-0.4}$ & $3.24^{+0.04}_{-0.05}$\\ 
$354.5+03.3$   &  $1.14\pm0.06$   & $0.2\pm{0.1}$  &  $3.9\pm0.7$ &  $4.0\pm{0.8}$ &  \dots  &  $0.4^{+0.1}_{-0.1}$ &  \dots   & $4.40^{+0.07}_{-0.06}$ \\ 
$355.4-02.4$   &  $1.28^{+0.03}_{-0.04}$   &  $2.5^{+0.7}_{-0}$ &  $7.1^{+0.7}_{-0.6}$ &  $8.0^{+0.7}_{-1.0}$ &  $8^{+1}_{-1}$ &  $1.4^{+0.2}_{-0.2}$ &  $3.4^{+0.5}_{-0.5}$ & $2.93\pm0.04$ \\ 
$355.9+03.6$   &  $0.62^{+0.02}_{-0.02}$   &  $1.7^{+18}_{-0}$ &  $0.52^{+0.06}_{-0.11}$ &  $0.74^{+17}_{-0}$ &  $0.6^{+0.8}_{-0}$ &  $0.46^{+0.07}_{-0.08}$ &  $0.49^{+0.08}_{-0}$ & $3.05\pm0.04$ \\ 
\hline
\end{tabular}
}
\end{table*}


The {\sc{Nebular Empirical Analysis Tool}} (NEAT, \citealp{Wesson2012}) was used to compute electron temperatures and densities as well as a variety of elemental and ionic abundances. NEAT uses multiple line iterations of line flux ratios to calculate the nebular parameters. The temperatures and densities used for the abundance calculations, as discussed in \citet{Wesson2012}, are those  most appropriate for their ionization potentials. The \citet{Kingsburgh1994} ionization correction factor (ICF) scheme is used to correct for unobserved ions. Extinction is compensated for using the \citet{Howarth1983} Galactic law. Logarithmic values of extinction were generally in the range 2.9 to 4.4 except for PNG 004.0-03.0 whose c(H$\beta$) value was 1.0. The atomic data used for collisionally excited lines are primarily that of CHIANTI 5.2 \citealp{Landi2006} with alternative parameters for O$^+$ and S$^{2+}$ (respective transition probabilites: \citealp{Zeippen1982, Mendoza1982}, respective collision strengths: \citealp{Pradhan1976, Mendoza1983}). Optical recombination line data is also used within NEAT, and the atomic data come from a variety of sources (\citealp{Davey2000, Kisielius1998, Liu1995, Storey1994, Pequignot1991, Escalante1990})- for a detailed breakdown, see Table 1 of \citet{Wesson2012}.

Not all lines listed in Table \ref{OptRes} are able to be used as input to NEAT. Those that appear both in Table \ref{OptRes} and in the accompanying line list for NEAT were used. Statistical uncertainties are calculated within NEAT using a Monte Carlo scheme, based upon the line flux uncertainties and their propagation through the diagnostics and abundance calculations - further information and a detailed analysis of error propagation within the code can be found in \citet{Wesson2012}. 

Table \ref{TeRes} lists the calculated electron temperatures and densities for a variety of ions in the UVES sample nebulae. The ionic and elemental abundances, ordered by element, are given in Table \ref{abns}. NEAT only gave solutions at exactly 20,000 K for T$_{\rm{e}}$(O III) in PNG $004.0-03.0$ and thus had zero statistical error so we considered these values to be unrealistic and unreliable and so did not include it in the table. The models may well have failed because of averaging of very different regions in this complex object.

\subsection{Comparisons with literature values}

\subsubsection{T$_{\rm{e}}$ and N$_{\rm{e}}$}

In general, our values of electron temperatures and densities are in good agreement with literature values to within the quoted uncertainties. Direct comparisons can be made in Table \ref{TeRes}, but the two cases where values derived in this work and those of the literature vary significantly will be discussed below.   

PNG $004.0-03.0$ had no literature value for T$_{\rm{e}}$(N II) so no direct comparison could be made. \citet{Exter2004} determined T$_{\rm{e}}$(O III) for this nebula, which we were unable to do, and found a value of $1.9\pm{0.3}\times 10^4$ K, which is $\sim8\times10^3$ K higher than our value for T$_{\rm{e}}$(N II) at $1.09\pm{0.04}\times10^4$ K. Our T$_{\rm{e}}$(N II) value is lower than what would be expected from the relation between T$_{\rm{e}}$(O III) and T$_{\rm{e}}$(N II). The lack of coincident determinations of T$_{\rm{e}}$(O III) and T$_{\rm{e}}$(N II) makes it impossible to determine whether the source of the difference is due to different regions being traced or whether there is a systematic difference in the calculations.

Our results for T$_{\rm{e}}$(O III) and T$_{\rm{e}}$(N II) in PNG $355.4-02.4$ agree to within the measured uncertainties of the values reported by \citet{Cuisinier2000} and our value of N$_{\rm{e}}$(O II) has reasonable agreement with their measured N$_{\rm{e}}$(S II). However, our  N$_{\rm{e}}$(Cl III) and N$_{\rm{e}}$(Ar IV) are higher by factors of 2-3, implying that the regions traced by Cl III and Ar IV are higher density than that of O II and S II or that the critical densities, and therefore the regimes of sensitivity to density, are different for the different indicators.

\subsubsection{Abundances}

In contrast to the relatively high levels of agreement between electron temperatures and densities in the literature and in this paper, the ionic and elemental abundances show some fairly significant differences. Our results from the aforementioned analysis are shown in Table \ref{abns}. Our results for PNG $004.0-03.0$ for all ions and elements lie between the results of \citet{Exter2004} and \citet{Miszalski2011}, with the exception of the elemental abundance of Ar for which our value is a factor of $\sim4$ smaller than the presented value of \citet{Exter2004} and a factor of 7 smaller than that of \citet{Miszalski2011}. The main origin of this difference is likely due to different line flux measurements in the observations as this nebula is known to possess a high density core which would affect the line flux ratios observed depending upon the exact pointing and slit widths used. Another potential source is the use of different ICFs; we use only [Ar IV] whereas other papers use both [Ar III] and [Ar IV].

Some nebulae, however, show reasonable agreement with a number of the literature values. Our calculated elemental and ionic abundances of helium, oxygen and nitrogen in PNG $006.1+08.3$ agree with those of \citet{Wang2007}. Additionally, S$^{2+}$/H$^+$ and S/H also agree well, but there is a discrepancy between the S$^{+}$/H$^{+}$ abundance: our value is nearly a factor of three lower than \citet{Wang2007}. However, as this ion has a very small impact on the overall abundance of sulphur, the two elemental abundances remain in good agreement. Abundances of argon, chlorine and neon are lower in our results than in those of \citet{Wang2007}: these vary between factors of 1.3 and 4.3, with the largest differences found in the Ar abundances (factor of three for Ar$^{3+}$/H$^{+}$ and a factor of four for Ar/H). These could be due to the different ICF schemes used or propagation of a single discrepant ion-abundance through the ICFs. Our elemental abundance results for PNG $355.4-02.4$ are generally in agreement or, at worst, within $2\sigma$ of the results of \citet{Cuisinier2000}. The only exception to this is N/H which is a factor of ten higher in our results than in \citet{Cuisinier2000} - this result is also higher than any of our sample, suggesting that this value is unreliable.

Only limited comparisons of our results to \citet{Gorny2004} can be made for PNG $003.6+03.1$ and PNG $354.5+03.3$. The results of He/H agree well for both nebulae. O$^{++}$/H$^+$ is higher in our results than \citet{Gorny2004} for both nebulae: our results are a factor of 1.8 higher in PNG $003.6+03.1$ and a factor of 3 in PNG $354.5+03.3$. Despite this, the overall O/H ratio in PNG $003.6+03.1$ agrees well with \citet{Gorny2004}: $4.4\times10^{-4}$ (this work) compared with $4.0\times10^{-4}$. Also, the O$^+/$H$^+$ ratio from our work in $354.5+03.3$ is consistent with that of \citet{Gorny2004}. The elemental abundance of nitrogen and sulphur are also higher in our results than \citet{Gorny2004} for PNG $003.6+03.1$ by factors of 2-3.

Three of our nebulae can be compared with the results of \citet{Gorny2009}: PNG $006.4+02.0$, PNG $006.8+04.1$ and PNG $355.9+03.6$. Our He/H values are consistently lower but the values lie within $2\sigma$ of those derived by \citet{Gorny2009}. The values of O/H and N/H in each nebula agree to within the calculated uncertainties, except for the N/H abundance in PNG $006.4+02.0$ which is $\sim2\sigma$ higher in our work. In contrast, the elemental abundances of argon and chlorine in PNG $006.4+02.0$ and PNG $006.8+04.1$ are a factor of ten lower in our work. \citet{Gorny2009} do not report the individual ionic abundances, so it is difficult to conclusively identify the origin of the discrepancy between our two works, but there are several potential sources. Firstly, \citet{Gorny2009} use the ICF scheme of \citet{Kingsburgh1994} for Ar and the scheme of \citet{Liu2000} for Cl, whereas we use the former for both Ar and Cl. This may be the root of some of the discrepancy in Cl abundances. With Ar, \citet{Gorny2009} observe, in addition to the Ar lines in this work, the [Ar III] line at 7135\AA, so the difference may be due to uncertainties in the ICF. Finally, the differences could also be due to differing atomic data used within ABELION (their work, atomic parameters from: \citealp{Galavis1998, Ramsbottom1997}; \citealp{Mendoza1982, Mendoza1983}) and NEAT (our work, atomic parameters from \citealp{Landi2006}), see \citet{Wesson2012} and \citet{Gorny2009} for further details.

\section{ISAAC infrared data}\label{ISAACdata}

\subsection{New Observations}

A total of 14 targets were observed at the Very Large Telescope in service mode between 2013-07-21 and 2013-07-30 under program ID 290.D-5136(A). Four of these were observed outside the requested observational constraints and, although the planetary nebulae had at least borderline detections of hydrogen, the spectra suffered from very poor signal-to-noise ratios and thus did not yield meaningful limits. A further four nebulae (PNG $006.3+04.4$, PNG $009.6-10.6$, PNG $351.9-01.9$ and PNG $353.5-04.9$) resulted in null detections within the requested observational constraints and are not discussed further. Of the remaining six nebulae, five were detected in the 3.625 $\upmu$m line and the final source had a high signal to noise ratio spectra with no detected [Zn IV] line, allowing useful upper limits to be placed on the Zn abundance within this nebula.

The observations used the Long Wavelength Spectrometer of ISAAC on UT3 using the jiggle-nod method, with on-source exposure times of 30-60 minutes. The slit length was $120\arcsec$, the slit width used was $1.5\arcsec$, the wavelength coverage was 3.55-3.80 $\upmu$m and the resulting resolution was $R=1500$. Nodding was done along the slit, with nods of 15$\arcsec$ to $30\arcsec$ depending upon the source being observed. This results in the source being in the field of view at all times during the observation, reducing the required observing time whilst still allowing an effective background subtraction. The slit was aligned through the brightest portion of the nebulae and in most cases included more than 50\% of the object due to their small angular size. The observed wavelength range covers the H I (n=8-5) line, five Humphreys series hydrogen lines (n=21-6 to n=17-6) and the [Zn IV] emission line at 3.625 $\upmu$m.

The spectra were reduced using the ESO common pipeline library. The wavelength calibration used the recommended method of arc lamp line calibration. The relative intensity calibration along the spectrum was applied using the telluric standard stars (details of which can be found in Table \ref{tellurics}), which are well described by black body distributions. We estimate the relative flux calibrations to be accurate to 5-10\%.  Absolute flux calibrations are not required for the analysis in this paper: the [Zn IV] line integrated intensities are given in terms of the H$_{8-5}$ line. Extinction is negligible over this wavelength range. 

\begin{figure*}
\centering
\subfigure{\includegraphics[trim=2cm 13cm 2cm 3cm, clip=true,width=0.45\textwidth ]{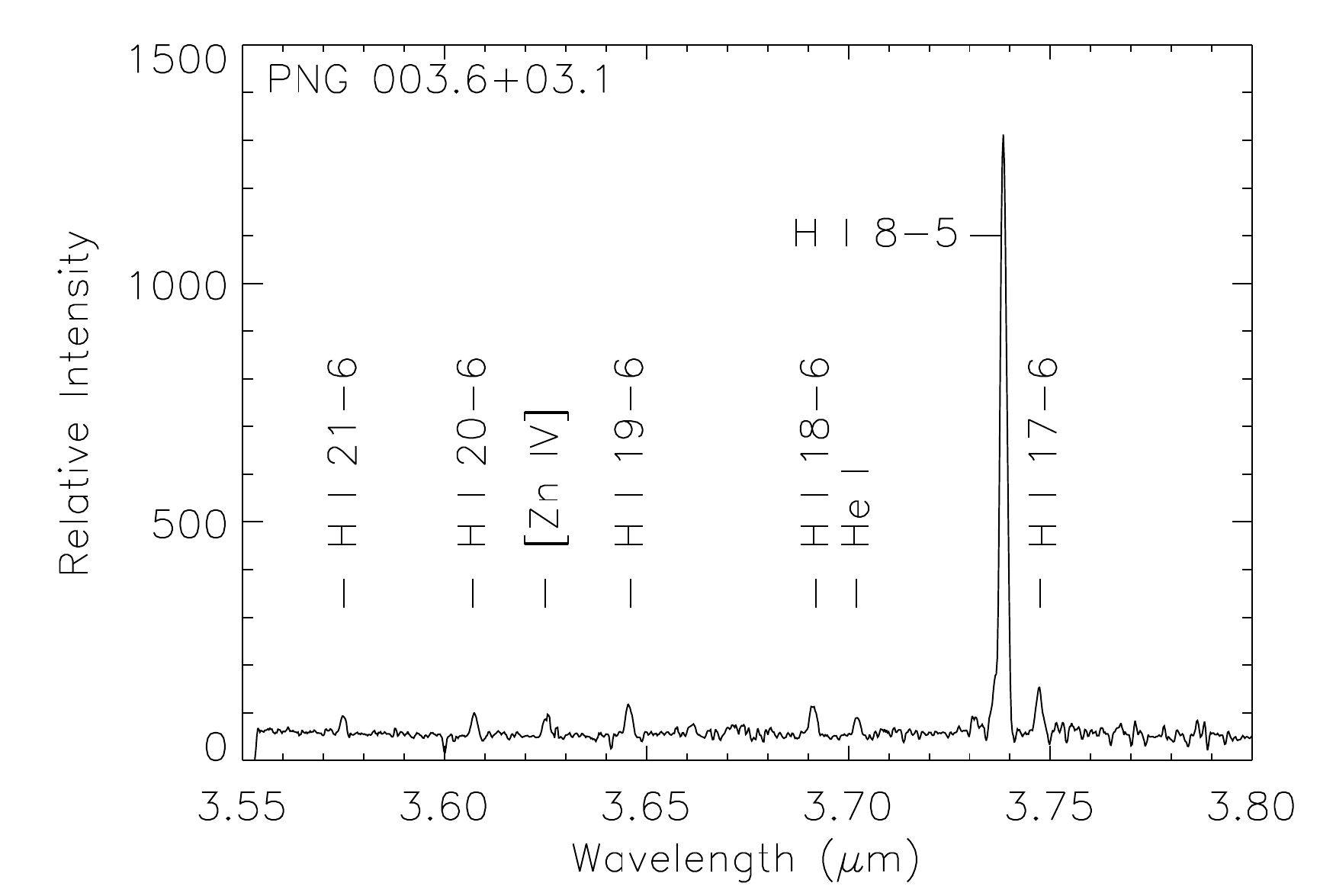}}
\subfigure{\includegraphics[trim=2cm 13cm 2cm 3cm, clip=true,width=0.45\textwidth]{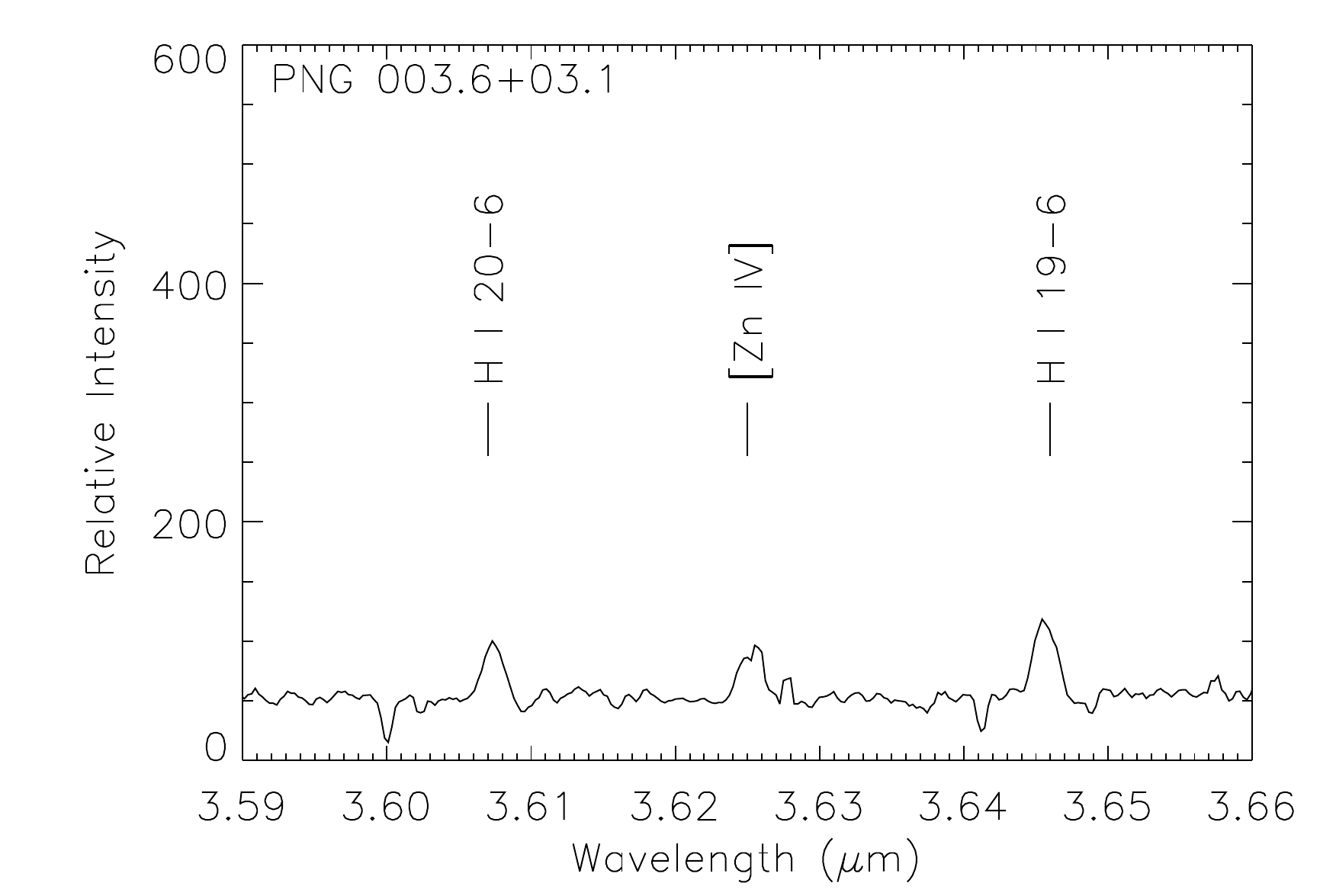}}
\subfigure{\includegraphics[trim=2cm 13cm 2cm 3cm, clip=true,width=0.45\textwidth]{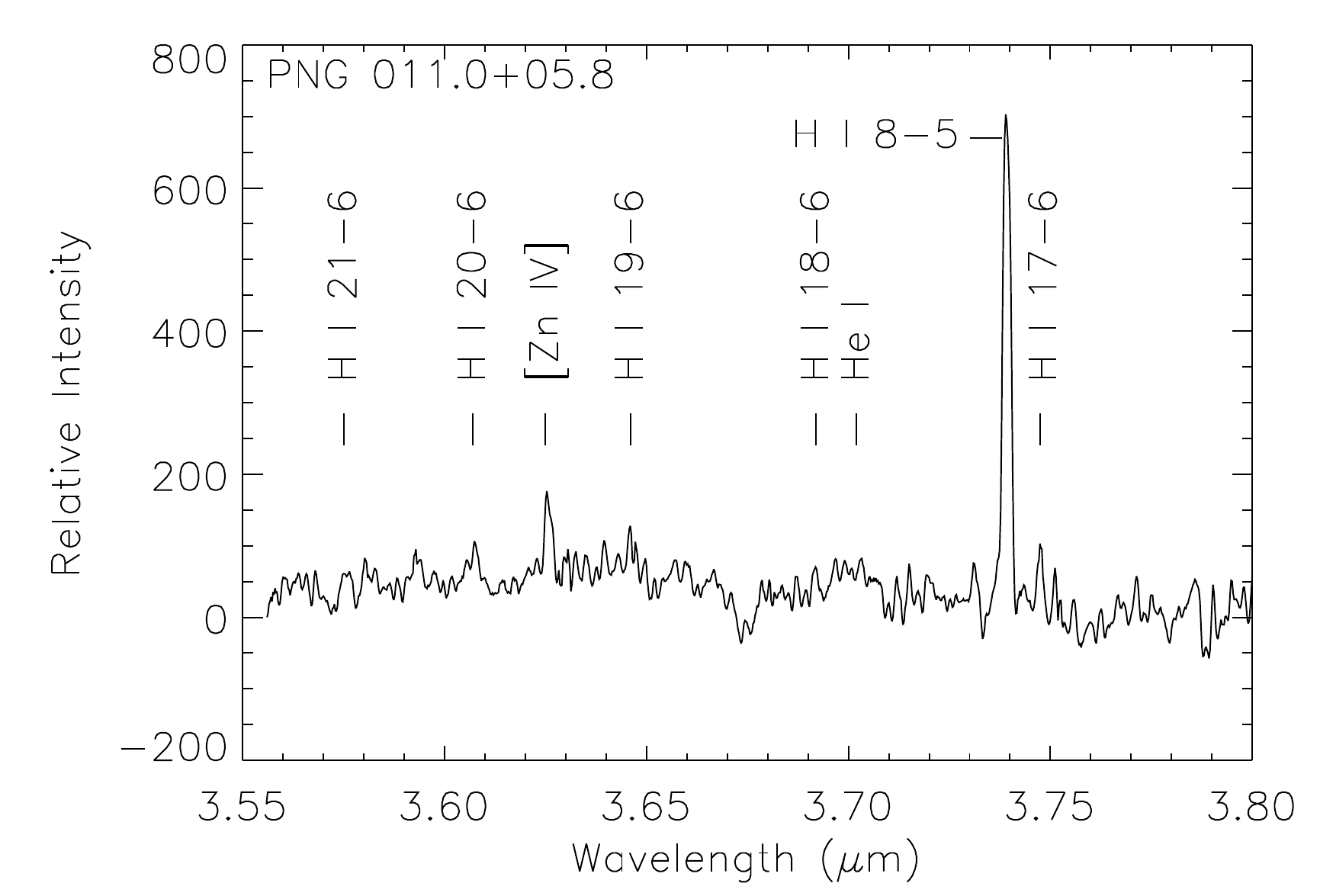}}
\subfigure{\includegraphics[trim=2cm 13cm 2cm 3cm, clip=true,width=0.45\textwidth]{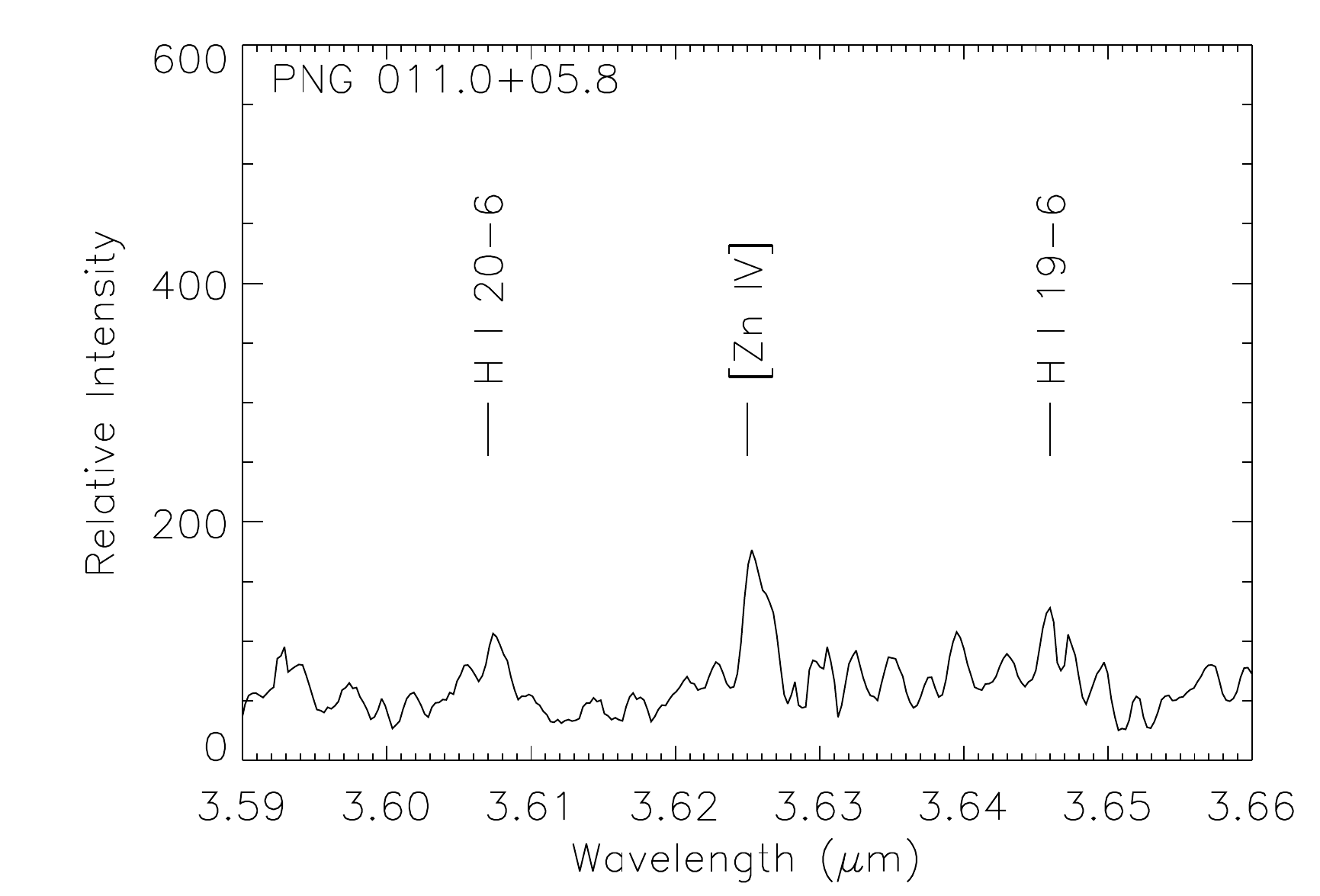}}
\subfigure{\includegraphics[trim=2cm 13cm 2cm 3cm, clip=true,width=0.45\textwidth]{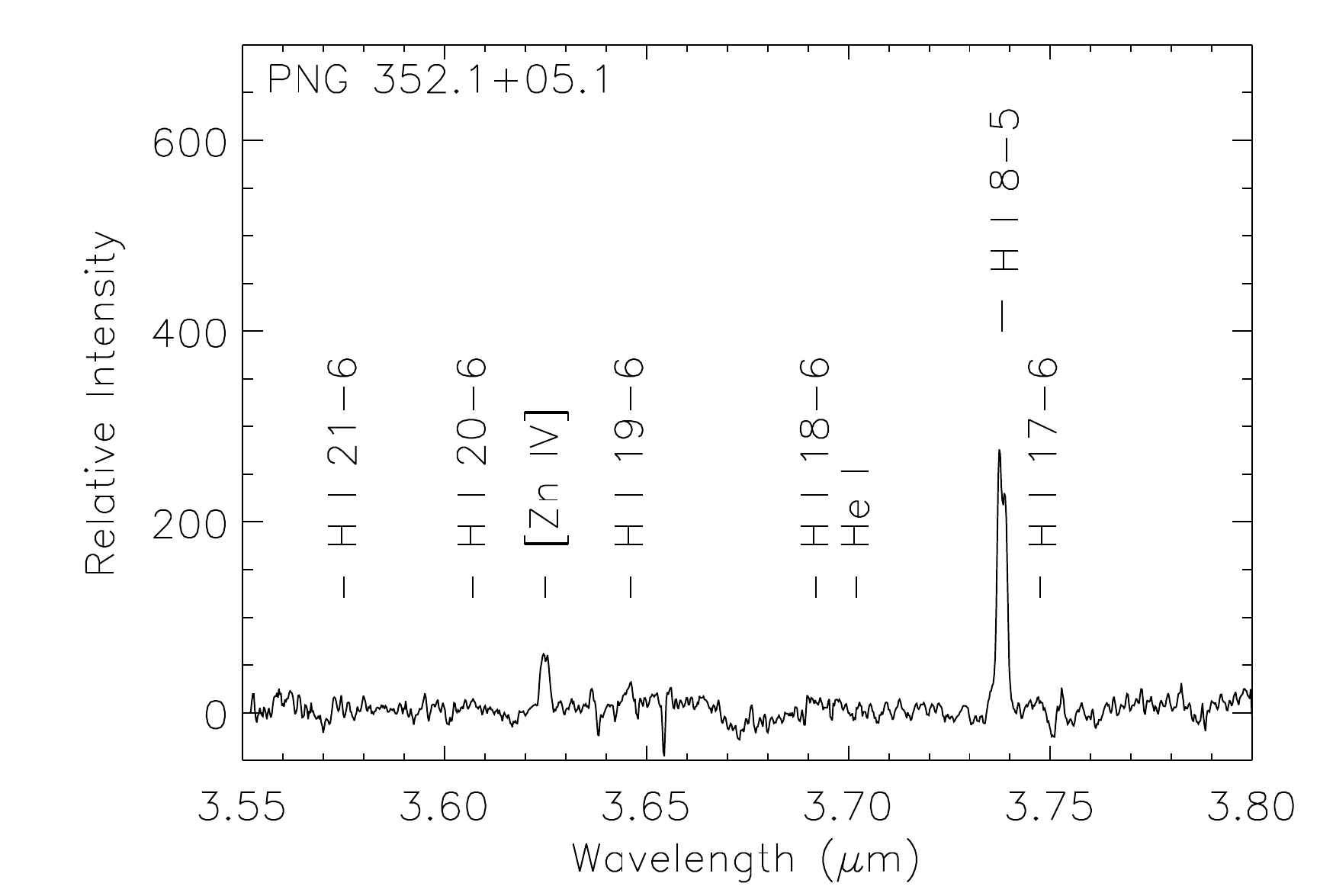}}
\subfigure{\includegraphics[trim=2cm 13cm 2cm 3cm, clip=true,width=0.45\textwidth]{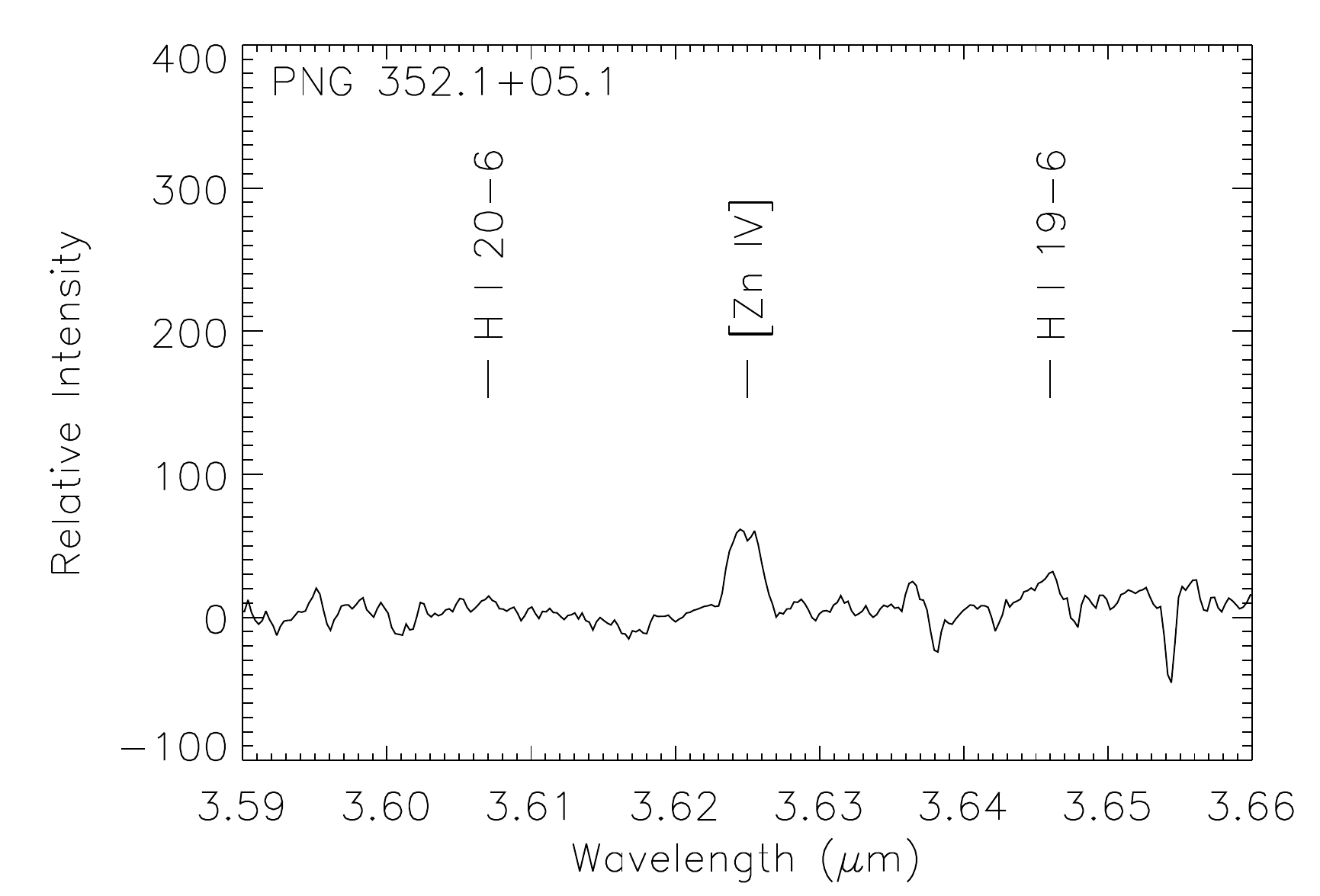}}
\subfigure{\includegraphics[trim=2cm 13cm 2cm 3cm, clip=true,width=0.45\textwidth]{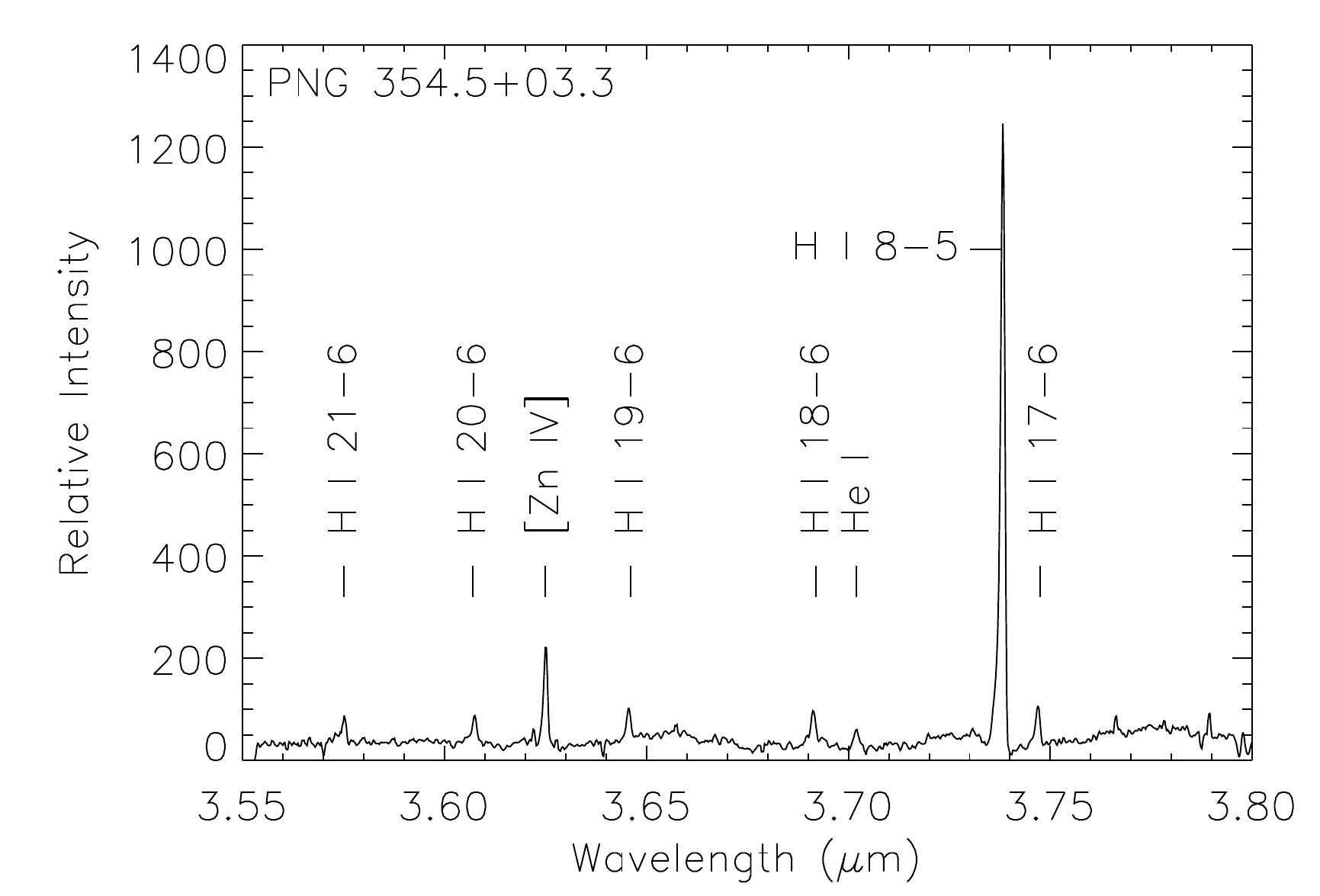}}
\subfigure{\includegraphics[trim=2cm 13cm 2cm 3cm, clip=true,width=0.45\textwidth]{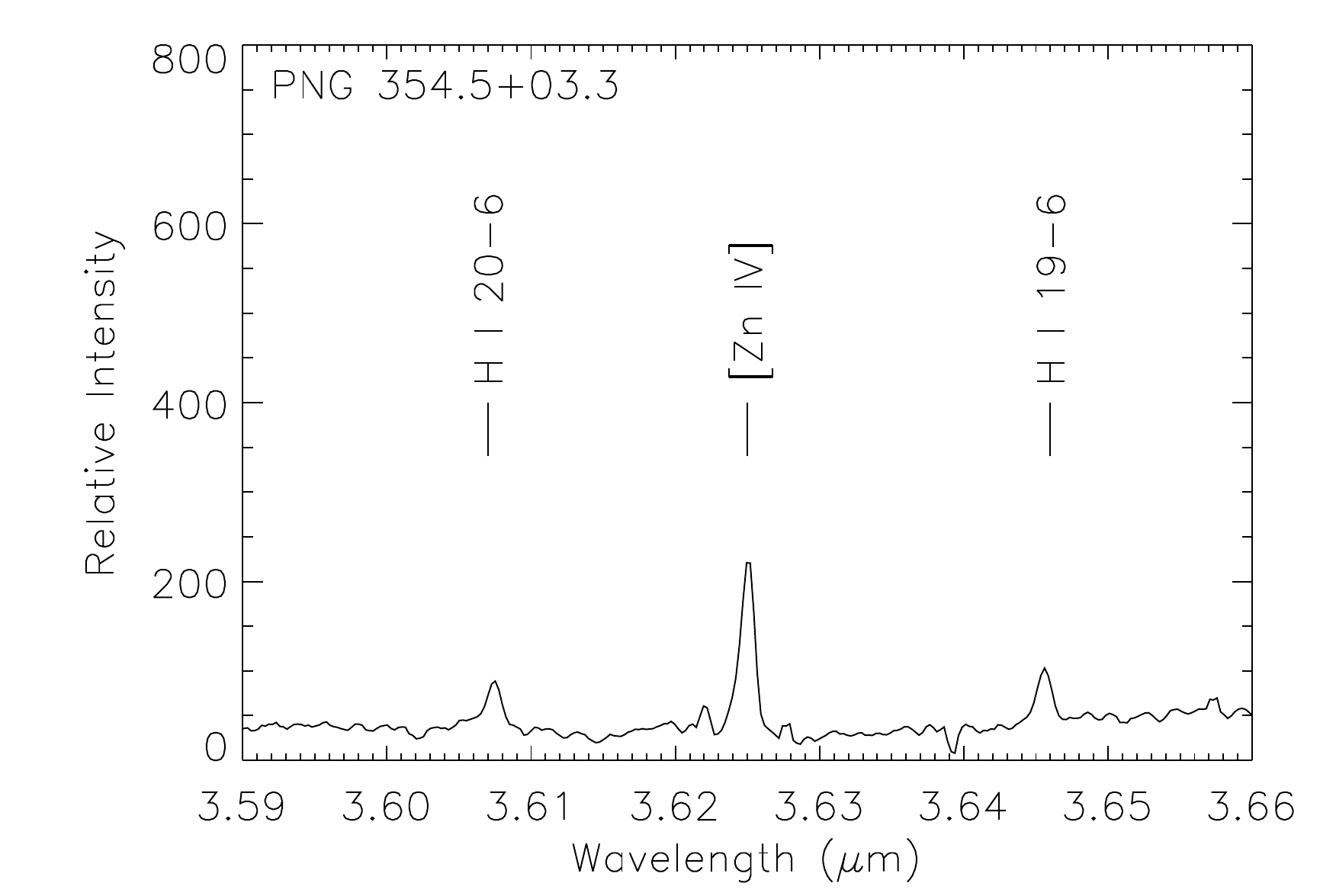}}
\caption{Spectra of all sources with detections or usable limits of the  [Zn IV]  line.  The spectra have been background subtracted and subsequently smoothed.}\label{spectra}
\end{figure*}
\begin{figure*}
\centering

\subfigure{\includegraphics[trim=2cm 13cm 2cm 3cm, clip=true,width=0.45\textwidth]{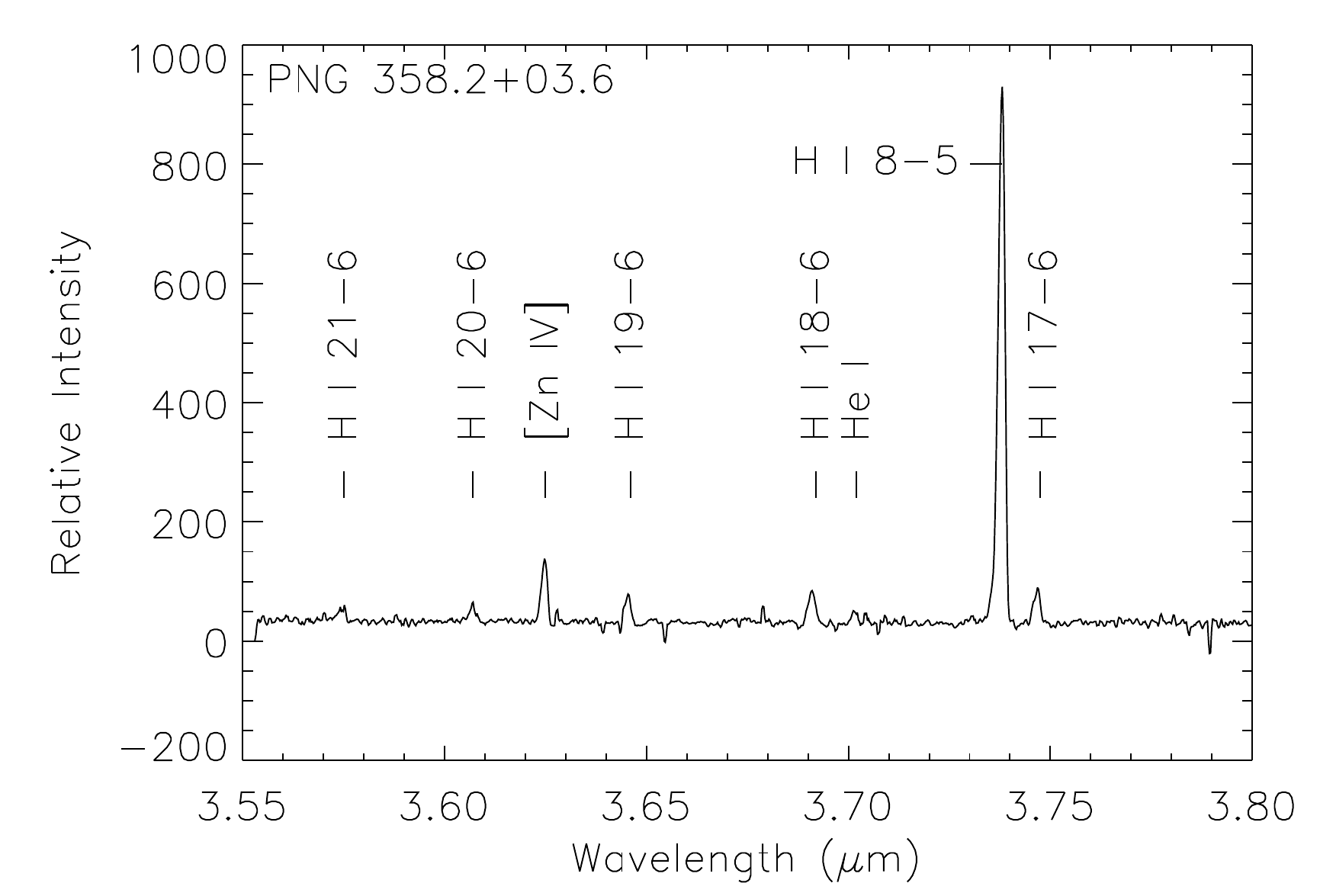}}
\subfigure{\includegraphics[trim=2cm 13cm 2cm 3cm, clip=true,width=0.45\textwidth]{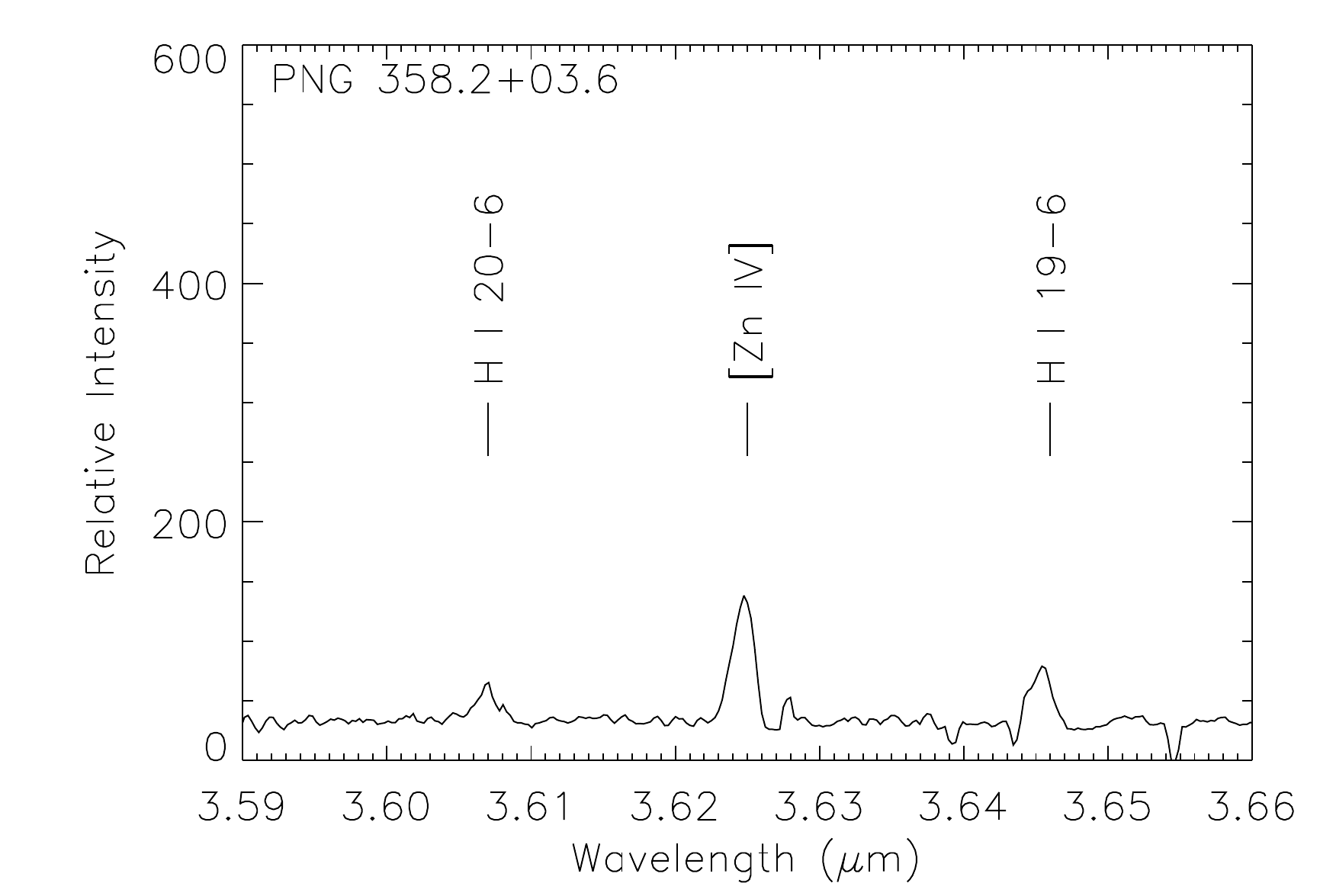}}

\subfigure{\includegraphics[trim=2cm 13cm 2cm 3cm, clip=true,width=0.45\textwidth]{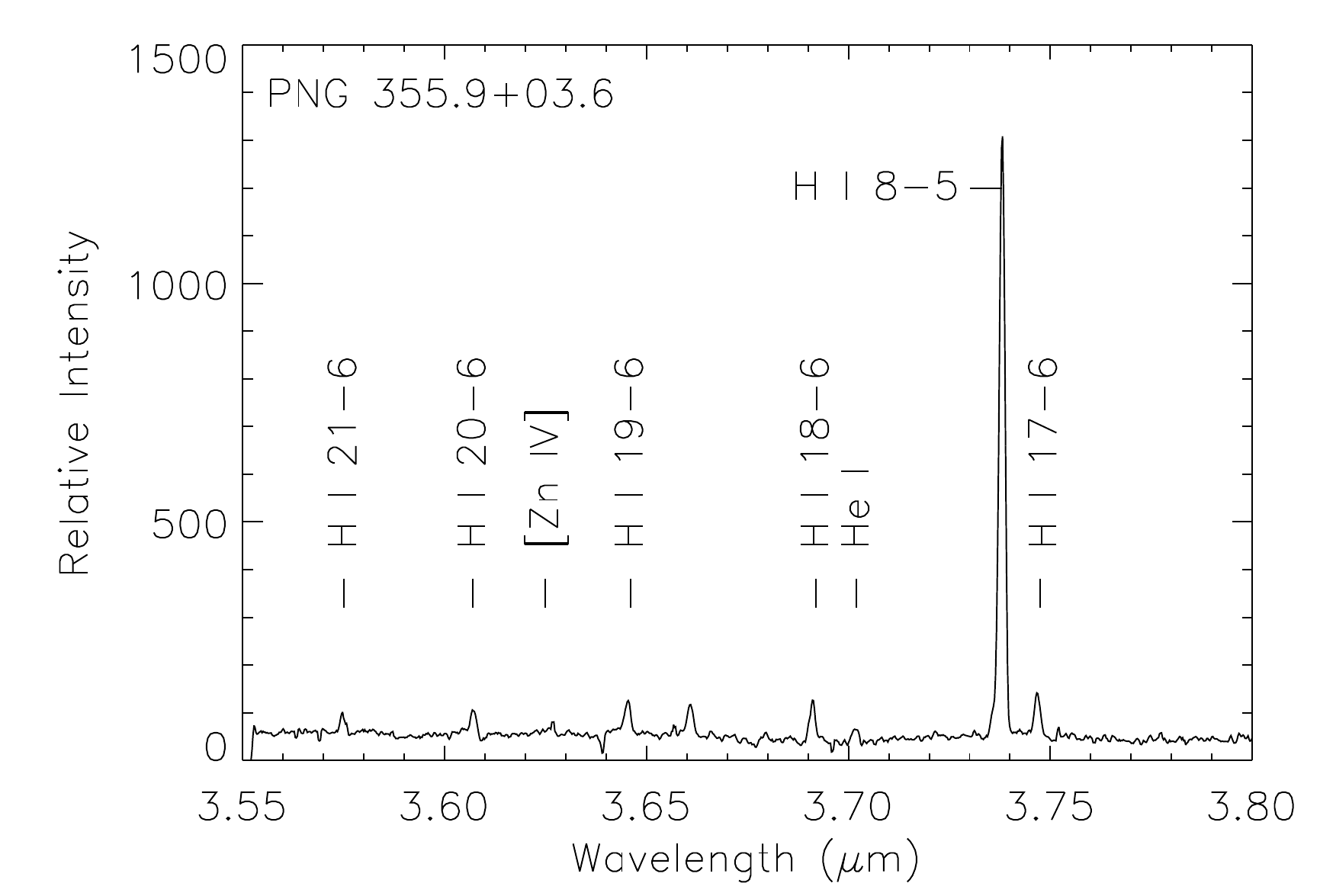}}
\subfigure{\includegraphics[trim=2cm 13cm 2cm 3cm, clip=true,width=0.45\textwidth]{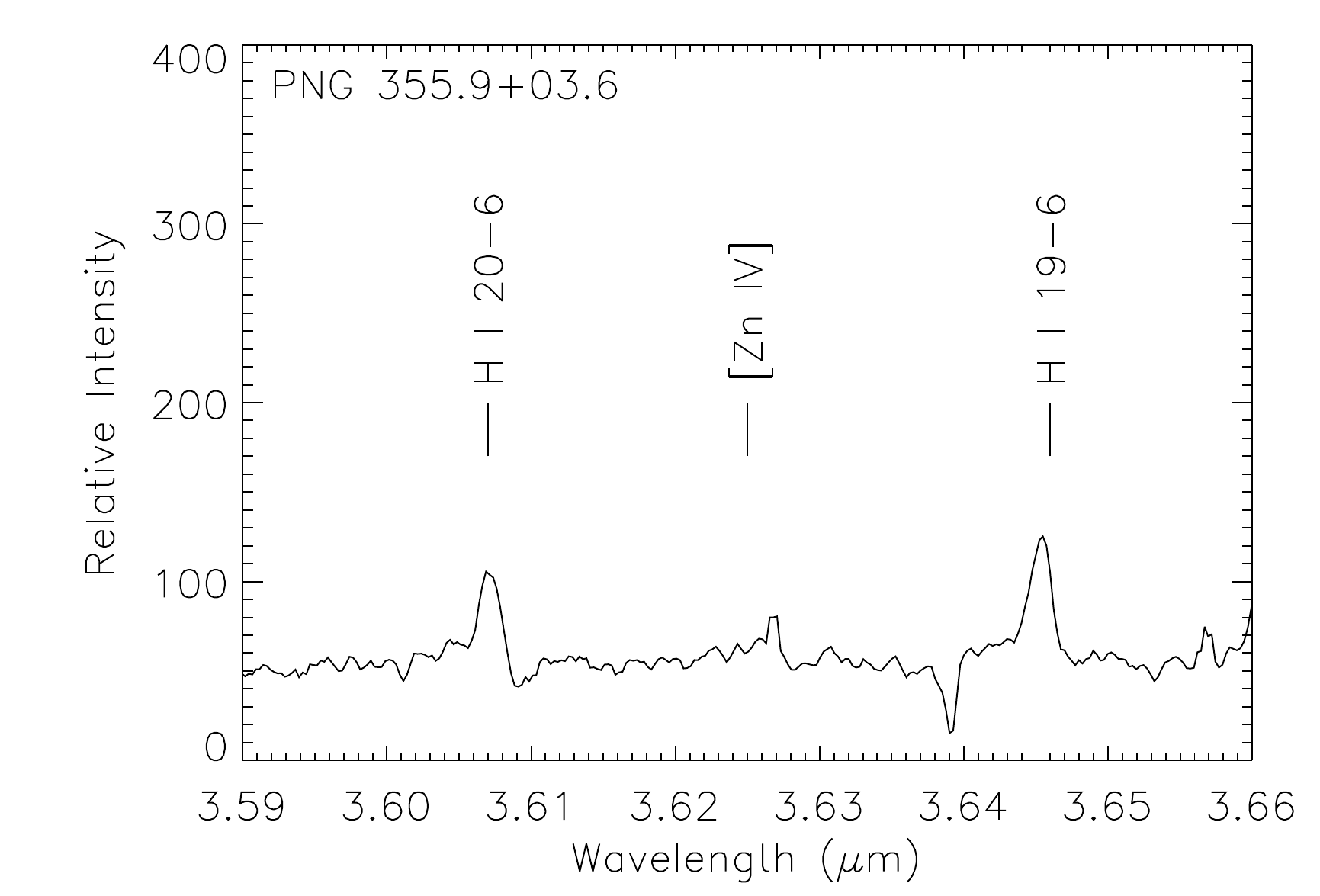}}
\contcaption{}
\end{figure*}

\begin{table}
\renewcommand{\arraystretch}{1.2}
\centering
\caption[Telluric standards for each source]{Telluric standards for each source, including their spectral type and effective temperatures, taken from the Hipparcos Catalogue.\smallskip}\label{tellurics}
\begin{tabular}{c l l l}
\hline
PNG	&		Telluric Standard	& Spectral Type & T$_\text{eff}$ (K) \\
\hline
$003.6+03.1$	&	HIP094378	&	B5V		&	15,200 \\
$011.0+05.8$	&	HIP091014	&	B2III/IV	&	20,300 \\
$352.1+05.1$	&	HIP085885	&	B2II		&	20,000 \\
$354.5+03.3$	&	HIP103571	&	G0V		&	\phantom{0}5,940 \\
$355.9+03.6$	&	HIP087164	&	B2II		&	20,000 \\
$358.2+03.6$	&	HR7236		&	B9Vn		&	10,500 \\
\hline
\end{tabular}
\end{table}

The resulting spectra are shown in Fig. \ref{spectra} at two different wavelength ranges so the intensity of both the [Zn IV] and H$_{8-5}$ lines can be clearly seen. The integrated intensities of the [Zn IV] and H$_{8-5}$ lines were measured after smoothing was applied and a background continuum fitted. The ratio of the integrated intensities can be found in Table \ref{abs_abn}.

\subsection{Literature data}

In addition to the new infrared observations, we also update the zinc abundances of those nebulae previously observed with ISAAC (program ID: 089.D-0084(A), see \citealp{Smith2014b} for complete details on data reduction and calibration), which also fall within our UVES sample in Sect. \ref{UVESdata}. Within this work, we take the published [Zn IV]/$\text{H}_{8-5}$ flux ratios and derive new ionic and elemental abundances based upon the electron temperatures and densities discussed in Sect. \ref{UVESdata}.

\subsection{Determination of zinc abundances}

The zinc abundances were calculated using a version of the method used by \citet{Dinerstein2001} with a different ionization correction factor as {\sc{Cloudy}} (\citealp{Ferland2013}, v13.01) photoionization models have shown that the Zn$^{3+}$/Zn ratio is best traced by O$^{++}$/O rather than the Ar$^{3+}$/Ar ratio used in \citet{Dinerstein2001} \citep{Smith2014b}. The {\sc{Cloudy}} models  assumed spherical symmetry and a Solar composition (as specified in the {\sc{Cloudy}} documentation, Hazy Part 1: Table 7.1 - values taken from \citealp{Grevesse1998, Holweger2001}; and \citealp{Prieto2001, Prieto2002}). 

Using this method, Zn/H is given by:

\begin{equation}
\frac{\text{Zn}}{\text{H}}=\frac{F(\text{[Zn IV]})}{F(\text{H}_{8-5})}\frac{\varepsilon(\text{H}_{8-5})}{\varepsilon(\text{Zn IV})}\frac{\text{O}}{\text{O}^{++}},
\end{equation}

\noindent where F(X) and $\varepsilon$(X) denote the flux and emissivity of line X respectively. Zn/O can be determined from Zn/H using the value of O/H either directly calculated within this paper or from the literature.

The hydrogen emissivities were taken from the catalogue of \citet{Hummer&Storey}, interpolated to the literature electron temperature and density of each nebula (see Table \ref{params}). The [Zn IV] emissivities were calculated in the low-density limit from:

\begin{equation}
\varepsilon (\text{Zn IV}) = \text{exp}\left(-\frac{\Delta E_{ul}}{{\rm{k}}T_{\rm{e}}}\right)\frac{{\rm{h}}\nu_{ul}\beta\Upsilon}{g_l\sqrt{T_{\rm{e}}}},
\end{equation}

\noindent with all values in cgs units. $\beta$ is defined as \mbox{$\sqrt{(2\pi\hbar^4)/(m_{e}^3{\rm{k}})}=8.629\times10^{-6}$} (cgs units) and $\Upsilon$, the energy-averaged collision strengths of the transition, were kindly provided by K. Butler (priv. comm.). The collision strengths for [Zn IV] are estimated to be accurate to 20\% (K. Butler, priv. comm.).

The electron temperatures, densities, oxygen abundances (O/H and O$^{++}$/O), central star temperatures (where available) and basic information about each nebula are shown in Table \ref{params}. Where possible, these have been taken from the UVES data described in Sect. \ref{UVESdata}. Where this was not possible, the parameters were taken from the literature and the references for these values are indicated in the table. Also included in this table, for comparison purposes, are the nebulae from \citet{Dinerstein2001}. 

\begin{table*}
\centering
\caption{Basic information and literature parameters for each source. Angular diameters are in arc seconds, where $^\text{opt}$ indicates values derived from optical observations and $^\text{rad}$ denotes values derived from radio observations \citep{Acker1992}. The first seven nebulae are those from \citet{Smith2014b}, the following six are the ISAAC observations and the remaining two nebulae are from \citet{Dinerstein2001}. References for optical spectra and  central star effective temperatures  are listed in column 11.\smallskip}\label{params}
\renewcommand{\arraystretch}{1.2}
{\tabcolsep=0.11cm
\begin{tabular}{c c c c c c c c c c c c}
\hline
\multirow{2}{*}{PNG} & \multirow{2}{*}{Name} & \multirow{2}{*}{RA} & \multirow{2}{*}{Dec} & Ang. dia. & T$_{\text{star}}$ & Bulge & T$_\text{e}$(O III) & N$_\text{e}$(O II) & O/H & \multirow{2}{*}{O$^{++}$/O} & \multirow{2}{*}{Ref.} \\
 					 & 						 & 					   & 					  & (arcsec)  & ($10^4$ K) 		& /Disk  & ($10^4$ K) 			& ($10^3$ cm$^{-3}$) 			   & ($10^{-4}$) 				 & & \\
\hline 
$004.0-03.0$	& M 2-29 	& 18 06 41 	& $-$26 54 56 	& \phantom{0}3.6$^\text{opt}$	& \phantom{0}7.6 	& B & $1.09\pm{0.04}^\text{b}$	&$3.0^{+0.3}_{-0.4}$&  $1.2\pm{0.1}$  		& $0.92^\pm{0.11}$ 			& 1,2	\\
$006.1+08.3$	& M 1-20 	& 17 28 58 	& $-$19 15 54 	& \phantom{0}1.9$^\text{rad}$	& \phantom{0}7.9 	& B & $0.95^{+0.01}_{-0.09}$	&  $10^{+2}_{-0}$	&  $3.6^{+0.2}_{-0.1}$	& $0.94^{+0.04}_{-0.08}$	& 1,2	\\
$006.4+02.0$	& M 1-31 	& 17 52 41 	& $-$22 21 57 	& \phantom{0}7.0$^\text{rad}$	& \phantom{0}5.8	& B & $0.74^{+0.02}_{-0.01}$	&  $7^{+4}_{-0}$	&  $8.1^{+1.0}_{-0.2}$	& $0.98^{+0.10}_{-0.15}$	& 1,2	\\
$006.8+04.1$	& M 3-15 	& 17 45 32 	& $-$20 58 02 	& \phantom{0}4.2$^\text{opt}$	& \phantom{0}7.9	& B &  $0.81^{+0.01}_{-0.02}$	&  $5^{+2}_{-0}$	&  $5.8\pm{0.7}$  		& $0.99^{+0.15}_{-0.12}$	& 1,2	\\
$019.7+03.2$	& M 3-25 	& 18 15 17 	& $-$10 10 09 	& \phantom{0}3.9$^\text{opt}$	& \phantom{0}5.2	& D &  $1.09\pm{0.03}^\text{b}$	&  $14\pm{1}$ 		&  $3.9\pm{0.4}$  		& $0.82\pm{0.11}$ 			& 3,4	\\
$040.4-03.1$	& K 3-30 	& 19 16 28 	& +05 13 19 	& \phantom{0}3.3$^\text{rad}$	&  \dots 		& D	& 1.0$^\text{a}$  			&  10.0$^\text{a}$  &  3.9$^\text{a}$  		&  0.80$^\text{a}$ 			& -  	\\
$355.4-02.4$	& M 3-14 	& 17 44 21 	& $-$34 06 41 	& \phantom{0}2.8$^\text{rad}$	& \phantom{0}7.9 	& B & $0.81^{+0.02}_{-0.03}$	& $4^{+1}_{-0}$ 	&  $8.0^{0.7}_{-1.0}$	& $0.89^{+0.14}_{-0.11}$	& 1,2	\\
$003.6+03.1$	& M 2-14 	& 17 41 57 	& $-$24 11 16	& \phantom{0}2.2$^\text{rad}$	& \phantom{0}4.4 	& B & $0.68\pm{0.02}$ 			& $9^{+4}_{-0}$		& $4.4^{+0.8}_{-1.1}$ 	& $0.80^{+0.23}_{-0.15}$	& 1,2	\\
$011.0+05.8$	& NGC 6439 	& 17 48 20	& $-$16 28 44	& \phantom{0}5.0$^\text{opt}$	& \dots 		& D	& $1.01\pm0.01$ 			& $3.7\pm{0.1}$ 	& $5.0^{+0.3}_{-0.1}$ 	& $0.84^{+0.06}_{-0.03}$	& 3,-	\\
$352.1+05.1$	& M 2-8		& 17 05 31 	& $-$32 32 08	& \phantom{0}4.2$^\text{opt}$	& 12.8 			& B	& $0.94\pm{0.01}$ 			& $5\pm{3}$ 	& $4.8^{+0.3}_{-0.2}$ 	& $0.77^{+0.05}_{-0.06}$ 	& 3,2	\\
$354.5+03.3$	& Th 3-4	& 17 18 52 	& $-$31 39 07 	& \phantom{0}0.0$^\text{opt}$	& \dots			& B	& $1.06^{+0.06}_{-0.05}$ 		& $17^{+1}_{-2}$	& $4.0^{+0.8}_{-0.6}$ 	& $0.98\pm{0.26}$ 		& 1,2	\\
$355.9+03.6$	& H 1-9		& 17 21 32 	& $-$30 20 49	& \phantom{0}0.7$^\text{rad}$	& \phantom{0}3.8 	& B & $1.00^{+0.03}_{-0.04}$ 	& $30^{+60}_{-0}$ 	& $0.74^{+17}_{-0}$ 	& $0.70^{+0.08}_{-16}$		& 1,5	\\
$358.2+03.6$	& M 3-10	& 17 27 20 	& $-$28 27 51	& \phantom{0}3.2$^\text{opt}$	& \phantom{0}9.3 	& B & $1.08\pm0.03$ 			&$7.5^{+2.0}_{-1.4}$& $5.0^{+0.3}_{-0.4}$ 	& $0.82^{+0.08}_{-0.14}$	& 3,2	\\
$084.9-03.4$	& NGC 7027 	& 21 07 02 	& +42 14 10 	&  14.0$^\text{opt}$ 			&  18.0 	& D		&  $1.25\pm{0.04}$  		&  $13^{+10}_{-3}$ 	&  $3.9\pm{0.5}$  		&  $0.67\pm{0.01}$ 			& 6,7	\\
$089.8-05.1$	& IC 5117 	& 21 32 31 	& +44 35 48 	& \phantom{0}1.5$^\text{rad}$	&  12.0 		& D	&  $1.25\pm{0.04}$ 			&  $16^{+11}_{-4}$ 	& $2.7\pm{0.3}$  		&  $0.94\pm{0.13}$ 			& 8,8 	\\
\hline
\end{tabular}
}
{\begin{flushleft}\small
1: This work;  2: \citet{Gesicki2007}; 3: \citet{Gorny2004}; 4: \citet{Kondratyeva2003}; 5: \citet{Gorny2009}; 6: \citet{Zhang2005};  7: \citet{Pottasch2010};   8: \citet{Hyung2001}.

a: adopted value; b: T$_e$(NII) used.
\end{flushleft}}
\end{table*}

\subsection{Results}

The calculated abundances of Zn/H and Zn/O are shown in Table \ref{abs_abn}. The reference Solar values were taken from \citet{Asplund2009}. The table includes results that are unchanged from \citet{Smith2014b} (PNG $019.7+03.2$, PNG $040.4-03.1$, NGC 7027 and IC 5117) - these are included for completeness only. The results from this extended sample of nebulae are generally sub-solar in Zn/H. The values of O/Zn are either consistent with the Solar value to within uncertainties (8 of the sample), or are enriched in O relative to Solar (6 of the sample, $>2\sigma$ from Solar).

\begin{table*}
\centering
\renewcommand{\arraystretch}{1.2}
\caption[Abundances of zinc and iron for the new and literature samples.]{Abundances of zinc and iron with respect to hydrogen and oxygen for the new sample and the literature sample. Iron abundances may be calculated from the zinc abundances using the Solar ratio of Zn/Fe.  Flux ratios are listed in column 3 and are given as the flux of the [Zn IV] line with respect to to the flux of the H line used, H$_{8-5}$ unless otherwise indicated.  Emissivities are quoted in erg s$^{-1}$ cm$^{-3}$.\smallskip}  \label{abs_abn}
\begin{tabular}{c c c c c c c c}
\hline
\multirow{2}{*}{PNG}	& $\varepsilon$([Zn IV])&	F$_{[\text{Zn IV}]}$/F$_\text{H}$ 	& $\text{Zn}^{3+}/\text{H}^+$& Zn/H 					& Zn/O					& \multirow{2}{*}{[Zn/H]} 	& \multirow{2}{*}{[O/Zn]} 	\\
						& ($\times10^{-21}$)	& ($\times10^{-2}$)					  	& ($\times10^{-8}$)			 & ($\times10^{-8}$)		& ($\times10^{-5}$)		& 			    			&							\\
\hline
$004.0-03.0$			&	$7.2\pm{1.5}$		&	$\phantom{0}4.4\pm{1.5}$			&	$0.73\pm{0.28}$			 &	$0.79\pm{0.32}$			&	$6.6\pm{2.7}$		&	$-0.7\pm{0.1}$			&	$0.1\pm{0.2}$			\\
$006.1+08.3$			&	$7.0\pm{1.4}$		&	$\phantom{0}2.6\pm{0.4}$			&	$0.51^{+0.14}_{-0.13}$	 &	$0.54^{+0.16}_{-0.13}$	&	$1.5\pm{0.4}$		&	$-0.8\pm{0.1}$			&	$0.7\pm{0.1}$			\\
$006.4+02.0$			&	$6.5\pm{1.3}$		&	$\phantom{0}8.1\pm{1.1}$			&	$2.3^{+0.6}_{-0.6}$		 &	$2.3^{+0.7}_{-0.6}$		&	$2.9\pm{0.8}$		&	$-0.2\pm{0.1}$			&	$0.4\pm{0.1}$			\\
$006.8+04.1$			&	$6.7\pm{1.3}$		&	$\phantom{0}5.6\pm{0.8}$			&	$1.4\pm{0.3}$			 &	$1.4\pm{0.4}$			&	$2.4\pm{0.7}$		&	$-0.4\pm{0.1}$			&	$0.5\pm{0.1}$			\\
$019.7+03.2$ 			& 	$7.2\pm{1.4}$		& 	$17.4\pm{2.5}$ 						& 	$2.8\pm{0.7}$ 			 & 	$3.5\pm{0.9}$ 			& 	$9\pm{3}$ 			& 	$-0.02\pm{0.10}$ 		& 	$-0.08\pm{0.10}$		\\
$040.4-03.1^{\text{a}}$	&   $7.2\pm{1.6}$		&	$10.0\pm{1.4}$ 						& 	$1.8\pm{0.8}$ 			 & 	$2\pm{1}$ 				& 	$6\pm{5}$ 			& 	$-0.2\pm{0.2}$			& 	$0.1\pm{0.3}$ 			\\
$355.4-02.4$			&	$6.7\pm{1.3}$		&	$24.4\pm{3.5}$						&	$6.1\pm{1.5}$			 &	$6.8\pm{1.9}$			&	$8.5\pm{2.6}$		&	$0.3\pm{0.1}$			&	$-0.1\pm{0.1}$			\\
$003.6+03.1$			&	$6.3\pm{1.3}$		&	$\phantom{0}3.0\pm{0.4}$			&	$0.99^{+0.25}_{-0.24}$	 &	$1.2^{+0.4}_{-0.5}$		&	$2.8\pm{1.1}$		&	$-0.5\pm{0.1}$			&	$0.4^{+0.2}_{-0.1}$		\\
$011.0+05.8$			&	$7.1\pm{1.4}$		&	$13.8\pm{1.9}$						&	$2.5\pm{0.6}$		 	 &	$3.0^{+0.7}_{-0.8}$		&	$6.0^{+1.5}_{-1.6}$	&	$-0.1\pm{0.1}$			&	$0.1\pm{0.1}$			\\
$352.1+05.1$			&	$7.0\pm{1.4}$		&	$18.8\pm{2.7}$						&	$3.7\pm{0.9}$			 &	$4.9^{+1.3}_{-1.2}$		&	$10.2\pm{2.7}$		&	$0.1\pm{0.1}$			&	$-0.1\pm{0.1}$			\\
$354.5+03.3$			&	$7.2\pm{1.4}$		&	$14.4\pm{2.0}$						&	$2.4\pm{1.5}$		 &	$2.5\pm1.7$		&	$6\pm4$	&	$-0.2\pm0.2$	&	$0.1\pm0.2$	\\
$355.9+03.6$			&	$7.1\pm{1.4}$		&	$<0.9\pm{0.1}$						&	$<0.16\pm{0.04}$		 &	$<0.23^{+5.24}_{-0.06}$	&	$<3\pm{70}$			&	$<-1.2^{+1.4}_{-0.1}$	&	$>0.4\pm{1.4}$			\\
$358.2+03.6$			&	$7.2\pm{1.4}$		&	$9.9\pm{1.4}$						&	$1.6\pm{0.4}$			 &	$2.0^{+0.6}_{-0.5}$		&	$4.0^{+1.3}_{-1.1}$	&	$-0.3\pm{0.1}$			&	$0.3\pm{0.1}$			\\
NGC 7027 				& 	$7.4\pm{1.5}$ 		& 	$149\pm{7}^\text{b}$ 				& 	$0.88\pm{0.18}$			 & 	$1.3\pm{0.3}$ 			& 	$3.4\pm{0.8}$ 		& 	$-0.44\pm{0.08}$		& 	$0.34\pm{0.09}$ 		\\
IC 5117 				& 	$7.4\pm{1.5}$ 		& 	$260\pm{50}^\text{b}$ 				& 	$1.7\pm{0.4}$	 		 & 	$1.8\pm{0.5}$ 			& 	$7\pm{2}$ 			& 	$-0.31\pm{0.10}$		& 	$0.06\pm{0.12}$ 		\\
\hline
\end{tabular}
{\begin{flushleft} \small
a: values are an estimate using average values for oxygen ratios in calculations where literature values were unavailable

b: flux are given for the H$_{19-6}$ transitions and abundances are the mean of those derived from the H$_{19-6}$ and H$_{20-6}$ transitions, taken from \citet{Dinerstein2001}.
\end{flushleft}}
\end{table*}

The [Zn IV] 3.625 $\upmu$m line was not detected in PNG 355.9+03.6, thus only limits could be placed by measuring the integrated intensity of the region where the [Zn IV] line should have been observed and comparing this to the measured hydrogen integrated intensity. This gives limits on the values of [Zn/H] and [O/Zn] to be $-1.2$ and 0.4 respectively. All hydrogen lines from the Humphreys and Pfund series within this wavelength range were detected. This nebula has a very low central star temperature of $3.8\times10^{4}$ K, and it is at this temperature that [Zn IV] loses its dominance according to the {\sc{Cloudy}} models of \citet{Smith2014b}. Thus, this nebula may not be as metal-poor as these limits suggest. Taking Zn$^{3+}$/Zn to be 0.45 (the result obtained from the {\sc{Cloudy}} models of \citet{Smith2014b} at this central star temperature), the limits on [Zn/H] and [O/Zn] reduce to $<-1.0$ and $>0.2$ respectively. These values suggest that this nebula may have a significantly lower abundance of zinc in comparison to the Sun.

PNG $011.0+05.8$, PNG $352.1+05.1$ and PNG $354.5+03.3$ are consistent with Solar for both Zn/H and O/Zn. PNG $003.6+03.1$ and PNG $358.2+03.6$ both show sub-Solar [Zn/H] and above-Solar values for [O/Zn], indicating an enhancement in O/Zn in these nebulae.

Those nebulae in the \citet{Smith2014b} sample which also formed part of the UVES sample in Sect. \ref{UVESdata} had their abundance calculations updated using the UVES temperatures, densities and oxygen abundances. The updated results agree well to within uncertainties of the results originally obtained in \citet{Smith2014b} which were calculated using parameters derived from optical data from a variety of sources in the literature. PNG $004.0-03.0$ remains significantly sub-solar in [Zn/H] at $-0.7\pm{0.1}$ dex with an [O/Zn] value which is consistent with Solar to within the measured uncertainties. PNG $006.1+08.3$, PNG $006.4+02.0$ and PNG $006.8+04.1$ remain sub-solar in [Zn/H] at $-0.8\pm{0.1}$, $-0.2\pm{0.1}$ and $-0.4\pm{0.1}$ respectively and enhanced in [O/Zn] with respect to Solar at $0.7\pm{0.1}$, $0.4\pm{0.1}$ and $0.5\pm{0.1}$. PNG $355.4-02.4$ remains enhanced in zinc over Solar with [Zn/H]=$0.3\pm{0.1}$ but its [O/Zn] value has changed by +0.1, dex bringing it in line (within uncertainties) with the Solar value. The remaining nebulae detailed in \citet{Smith2014b} had no new optical data and thus their values remain unchanged and are included in this paper for completeness only.

\section{Discussion}

\subsection{Alpha element abundances as a function of metallicity}

\begin{figure}
\includegraphics[trim=0cm 0cm 0cm 16cm, clip=true,width=0.48\textwidth]{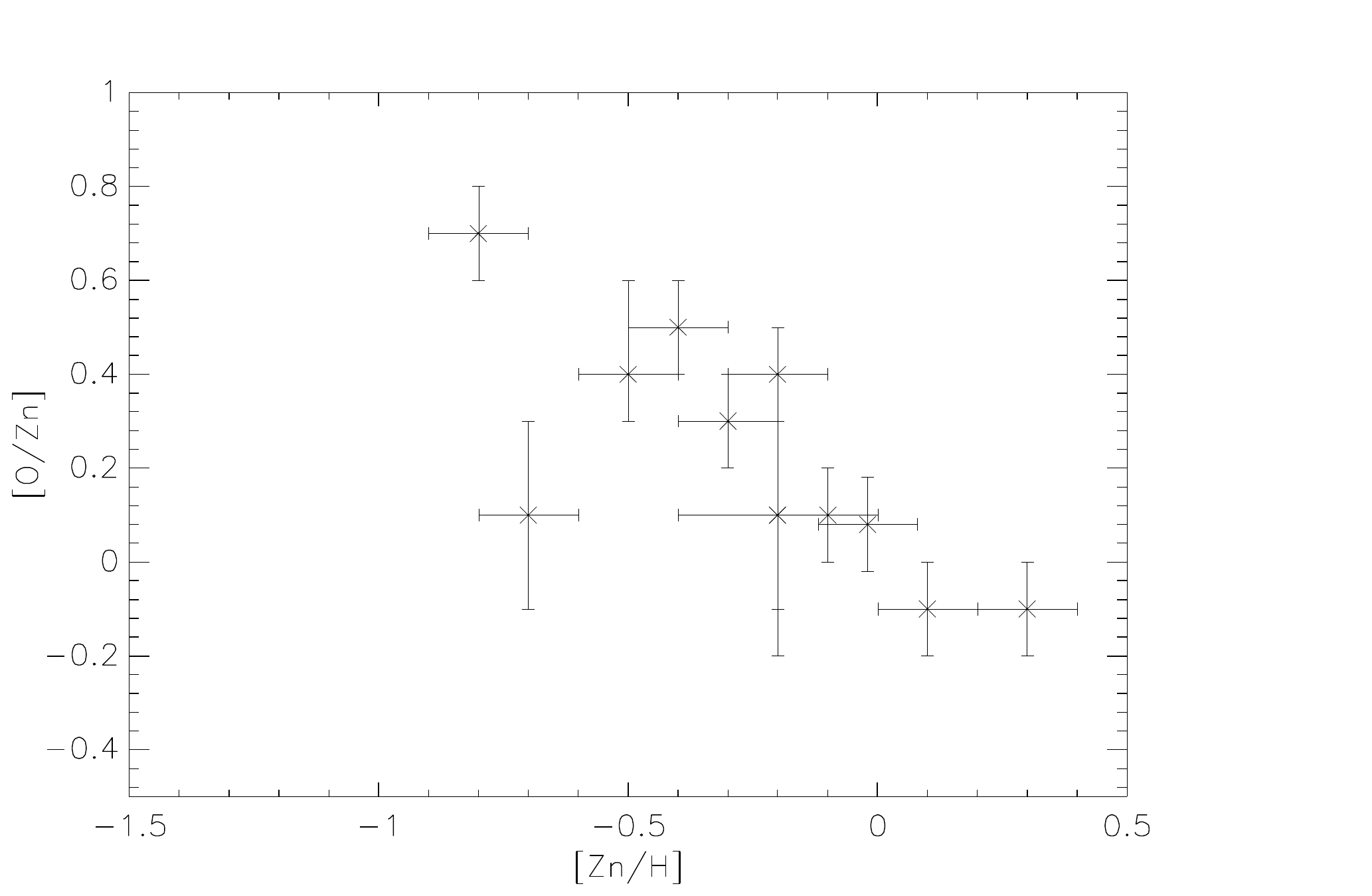}
\caption{[O/Zn] against [Zn/H].}\label{ZnH_OZn}
\end{figure}

Fig. \ref{ZnH_OZn} shows the relation between [O/Zn] and [Zn/H] in our sample nebulae. The two values are related by their O/H value. The majority of the nebulae follow the trend of decreasing [O/Zn] from 0.6 to -0.1 with increasing [Zn/H] from -1 to 0. Only limits were placed on PNG $355.9+03.6$ and thus it was not included in this figure. PNG $004.0-03.0$ lies approximately $2\sigma$ away from the remainder of the sample, due to its low [Zn/H] coupled with an approximately Solar [O/Zn]. Chemical evolution model results of \citet{Kobayashi2011} for the bulge show a decrease in [O/Fe] from 0.5 to 0 with [Fe/H] increasing from -1 to 0 as do the results of \citet{Mendelez2008} from observed abundances of Galactic Bulge stars. \citet{Bensby2014} and \citet{Jonsson2017} find that [O/Fe] decreases with increasing [Fe/H] from [Fe/H] = -0.75 to 0, reaching solar [O/Fe] at solar [Fe/H]. Our results are in excellent agreement with all of the aforementioned studies. The results of \citet{Schultheis2017} also show a strong negative correlation in [O/Fe] vs [Fe/H], with their results decreasing from [O/Fe]$\approx0.3$ at [Fe/H]$\approx-0.75$ to Solar values of [O/Fe] at [Fe/H] $\sim0.5$. This is a slightly shallower gradient than the previous studies, but is still generally consistent with the results of our work. Other alpha elements, such as Si, Ca and Mg, also in this study show similar trends to that of oxygen. The results of \citet{Barbuy2015}, however, show no trend in [O/Zn] with varying [Fe/H] (+0.3 to -1.3) or [O/H] (+0.6 to -0.4) in a sample of 56 Galactic bulge stars, which is inconsistent with our results, but the strong decrease in [O/Fe] with [Fe/H] in the sample of Bulge dwarf and sub-giant stars of \citet{Bensby2011} is in good agreement with our results. This agreement between our results and the literature studies indicates that the PNe have a similar star formation history to the stars in the optical studies.

In general, our sample have low [Zn/H] and solar [O/H] values implying that, assuming zinc abundances reflect those of iron and there is no depletion of zinc, there is some alpha-element enhancement in our sample. \citet{Bensby2017} report a range of metallicities and ages in the Bulge, including younger objects of Solar metallicity or higher. \citet{Gesicki2014} find that Bulge planetary nebulae are related to the younger population of the Bulge, which would fit with our data only if the zinc abundances were underestimated, resulting in Solar metallicity PNe with no alpha element enhancement. \citet{Zoccali2017}, however, report two bulge populations: one with super-solar metallicities and one with sub-solar metallicities. The zinc abundance results of our sample nebulae directly place them in the latter category, although their oxygen abundances are closer to Solar. Neither our oxygen nor zinc abundances could place our sample within the super-solar category.

\subsection{Optical abundances as a function of metallicity}

\begin{figure}
\includegraphics[trim=0cm 0cm 2cm 16cm, clip=true,width=0.48\textwidth]{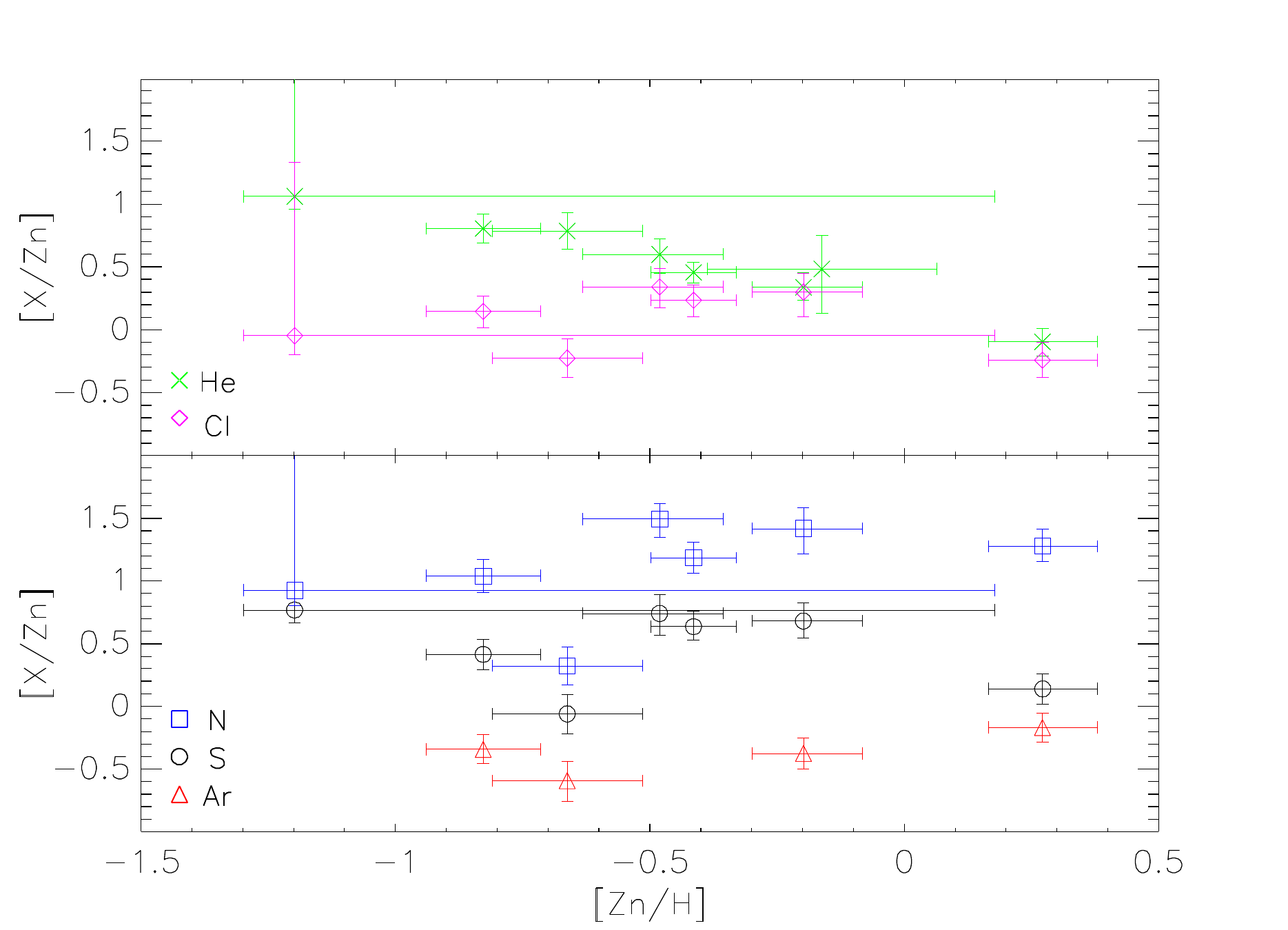}
\caption{[He/Zn], [N/Zn], [S/Zn], [Ar/Zn] and [Cl/Zn] abundances against [Zn/H]. Symbols are defined in the legends of each plot and the Solar values have been taken from \citet{Asplund2009}.}\label{opt_comp}
\end{figure}

The elemental abundances determined solely from the UVES data are shown in Fig. \ref{opt_comp} as a fraction of Zn, obtained by dividing the elemental abundances with respect to hydrogen by Zn/H, and are plotted as a function of [Zn/H].  The [X/Zn] ratios are generally constant or show slight decreases with increasing metallicity.

The chemical evolution models of \citet{Kobayashi2011} for the bulge show the variation of [Cl/Fe], [Ar/Fe], [N/Fe] and [S/Fe], amongst other elements, over $-1.5<$[Fe/H]$<0.5$. [S/Fe] decreases from 0.5 to 0 as [Fe/H] increases from -1 to 0. Our results show no clear decrease but are of similar values, distributed between 0.7 and -0.1. The lack of clear trend may be due to the small sample size, as observational data presented with the model results for the bulge are also subject to large scatter. [Cl/Fe] increases then decreases in the models of \citet{Kobayashi2011}, beginning at a value of -0.3, increasing to 0.1 then decreasing to -0.1. Our results are consistent with this model. [Ar/Fe] decreases from 0.3 to -0.1 over the same [Fe/H] range as the two aforementioned models. Our results are lower than these values and approximately constant, with a typical value of [Ar/Zn]=-0.5. The model [N/Fe] increases from -0.5 to 0 then decreases back to -0.3 as [Fe/H] increases from -1 to -0.5 and from -0.5 to 0 respectively. Our results show an approximately constant value of [N/Zn] with a significantly higher average value of 1.0.

\section{Summary and conclusions}

We have used UVES observations of planetary nebulae in combination with the {\sc{Nebular Empirical Analysis Tool}} to calculate nebular diagnostics such as electron temperatures, densities and ionic and elemental abundances. These values have been compared with literature data and in many cases agree well with previously published values. The electron temperatures and densities as well as the oxygen elemental and ionic abundances have been used in the analysis of VLT ISAAC observations of the 3.625 $\upmu$m [Zn IV] emission line, based upon the method of \citet{Dinerstein2001}. Six new VLT ISAAC observations have been obtained and the remaining data originated from \citet{Smith2014b}.

The new results show sub-solar [Zn/H] abundances and a range in [O/Zn] (+0.3 dex to -0.2 dex, excluding PNG 355.9+03.6). One nebula, PNG 355.9+03.6, had no detection of the [Zn IV] emission line but an upper limit was placed on [Zn/H], which puts this nebula at substantially sub-solar in [Zn/H]. Our results, taken in combination with those reported in \citet{Smith2014b}, indicate that the metallicity measured via zinc abundances of planetary nebulae in the Galactic bulge is generally sub-Solar and in a small number of nebulae, substantially sub-Solar ([Zn/H] $\leqslant -1.0$). 

Additionally, abundances of lighter elements, specifically sulphur, nitrogen, argon, helium and chlorine have also been examined as functions of [Zn/H]. [X/Zn], where X are the aforementioned elements, remains constant or shows slight decreases with increasing [Zn/H].

\section{Acknowledgements}

We thank Dr Keith Butler for providing information on the [Zn IV] line emissivity prior to publication and the ESO staff at Paranal for support whilst observing. Based on observations made with ESO telescopes at the La Silla Paranal Observatory under programme IDs: 075.D-0104, 089.D-0084(A) and 290.D-5136(A). This research was supported by the Science and Technologies Funding Council (STFC, UK) under grant ST/L000768/1. C.L.S. was supported by an STFC studentship and is now supported by a fellowship under the Integrating Atmospheric Chemistry and Physics from the Earth to Space (IACPES) Collaborative Research and Training Experience (CREATE) programme of NSERC. H.L.D. was supported by (US) grant NSF AST-0708425.

\bibliography{Zn_refs}

\end{document}